\documentclass[aps,prc,twocolumn,superscriptaddress,nofootinbib,longbibliography,floatfix,10pt]{revtex4-1}

\usepackage{graphicx}
\usepackage{dcolumn}
\usepackage{bm}
\usepackage{morefloats}
\usepackage{multirow}
\usepackage{amssymb}
\usepackage{amsmath}
\usepackage{xcolor}
\usepackage{fix-cm}
\usepackage{mathptmx} 
\usepackage[T1]{fontenc}
\usepackage[colorlinks,allcolors=blue]{hyperref}
\makeatletter
\def\NAT@def@citea{\def\@citea{\NAT@separator}}
\makeatother

\begin{document}

\title{Multi-nucleon transfer in ${}^{58}\text{Ni}+{}^{60}\text{Ni}$ and ${}^{60} \text{Ni}+{}^{60}\text{Ni}$ in stochastic mean-field approach}
\author{B. Yilmaz}
\affiliation{Physics Department, Faculty of Sciences, Ankara University, 06100 Ankara, Turkey}
\author{S. Ayik}
\affiliation{Physics Department, Tennessee Technological University, Cookeville, TN 38505, USA}
\author{O. Yilmaz}
\affiliation{Physics Department, Middle East Technical University, 06800 Ankara, Turkey}
\author{A. S. Umar}
\affiliation{Department of Physics and Astronomy, Vanderbilt University, Nashville, TN 37235, USA}

\date{\today}

\begin{abstract}
The multi-nucleon exchange mechanism in ${}^{58} \text{Ni}+{}^{60} \text{Ni}$ and ${}^{60}
\text{Ni}+{}^{60}\text{Ni}$ collisions is analyzed in the framework of the stochastic mean-field
approach. The results of calculations are compared with the TDRPA calculations and the recent data
of ${}^{58} \text{Ni}+{}^{60} \text{Ni}$. A good description of the data and a relatively good
agreement with the TDRPA calculations are found. 
\end{abstract}


\maketitle

The transfer of particles between two reacting nuclei is believed to have a profound impact on the
outcome of nuclear reactions. These include the observed reduction in the number of evaporated
neutrons from a compound nucleus linked to the excitation of the pre-compound collective dipole
mode~\cite{chomaz1993,dasso1995,flibotte1996,baran2001,simenel2001}, which is likely to occur when
ions have significantly different N/Z ratio, the influence of transfer on fusion particularly at
deep subbarrier 
energies~\cite{back2014,broglia1983,esbensen1989b,morton1994,timmers1998,stefanini2007,zagrebaev2003,zagrebaev2007b,umar2008a,corradi2009,zhang2010,kohley2011,montagnoli2013,gautam2014,bourgin2014,liang2016}
and the distribution of fragments in deep-inelastic and quasifission 
reactions~\cite{volkov1978,antonenko1991,adamian2003,aritomo2009,umar2016}. These reaction aspects are intimately related to the
dissipation and equilibration during the early stages of the
collision~\cite{evers2011,washiyama2009,oberacker2014,umar2017,wen2018} and at low energies also depend on the shell structure of
the participating nuclei~\cite{adamian1996,aritomo2006b,zagrebaev2007,kozulin2014b,mohanto2018} and is sensitive
to the details of the evolution of the shape of the composite system~\cite{umar2015a}. This is in
contrast to most classical pictures, which generally assume near instantaneous, isotropic
equilibration. For these low-energy heavy-ion collisions the relative motion of the centers of the
two nuclei is characterized by a short wavelength and thus allows for a classical treatment, whereas
the wavelength for the particle motion is not small compared to nuclear sizes and should be treated
quantum mechanically. The mean-field approach such as the time-dependent Hartree-Fock (TDHF)
theory~\cite{simenel2012,simenel2018} and its extensions provide a microscopic basis for describing the
heavy-ion reaction mechanism at low bombarding energies, and have been extensively used to study
particle
transfer~\cite{umar2008a,simenel2010,simenel2011,scamps2013a,scamps2015d,scamps2017b,sekizawa2013,sekizawa2014,sekizawa2016,sekizawa2017a}.

In recent work, Williams et al. presented an experimental investigation of the nucleon transfer in
the ${}^{58} \text{Ni}+{}^{60} \text{Ni}$ collisions at center of mass energies in the vicinity of
the fusion barrier~\cite{williams2018}. They analyzed the experimental data in conjunction with the numerical
simulations using TDHF theory and its extension, the time-dependent random-phase
approximation (TDRPA)~\cite{balian1984,broomfield2008,simenel2011}. At low energies, the TDHF provides a good
description of the mean evolution of the nuclear collective motion but fails to describe the
fluctuating dynamics of the collective motion. The authors, employing the Balian and V\'en\'eroni
formula (it is referred to as TDRPA in~\cite{williams2018}), calculated the dispersion of the
primary fragment distributions, and obtained a very good agreement with the experimental results.
Apparently, due to a technical difficulty of the approach, the authors interpret the experimental data
of the ${}^{58} \text{Ni}+{}^{60} \text{Ni}$ collisions, with the result of the calculations of the
symmetric ${}^{60} \text{Ni}+{}^{60} \text{Ni}$ collision at the same $E_\mathrm{c.m.}/V_B$ value,
where $V_B$ is the corresponding barrier height.

Here, we undertake a study for the same experimental data for ${}^{58} \text{Ni}+{}^{60} \text{Ni}$ system
by employing the stochastic mean-field (SMF) approach~\cite{ayik2008}. The SMF approach goes beyond
the mean-field approximation by incorporating the mean-field fluctuations into the description. The approach
relies on an ensemble of mean-field events specified with quantal and thermal fluctuations at the
initial state. It is possible to project the SMF on macroscopic variables which provide a much
easier description of the dissipation and fluctuation mechanism in terms of the relevant macroscopic
variables. The relevant macroscopic variables evolve according to the generalized Langevin
description characterized by a set of quantal transport coefficients. As described in
Refs.~\cite{ayik2017,ayik2018}, the transport coefficients are determined entirely by the occupied
TDHF wave functions. They include quantal effects due to the shell structure, contain the full
collision geometry and involve no adjustable parameters. Here, we perform calculations for the
nucleon exchange mechanism for range of impact parameters leading to
deep-inelastic collisions. In this case, due to the di-nuclear configuration of the collision, 
the relevant macroscopic variables are the number of
neutron and protons on either side of the window plane. Since ${}^{58} \text{Ni}$ is deformed, we carry
out calculations for side and tip 
\begin{figure}[!hpt]
\includegraphics*[width=8.5cm]{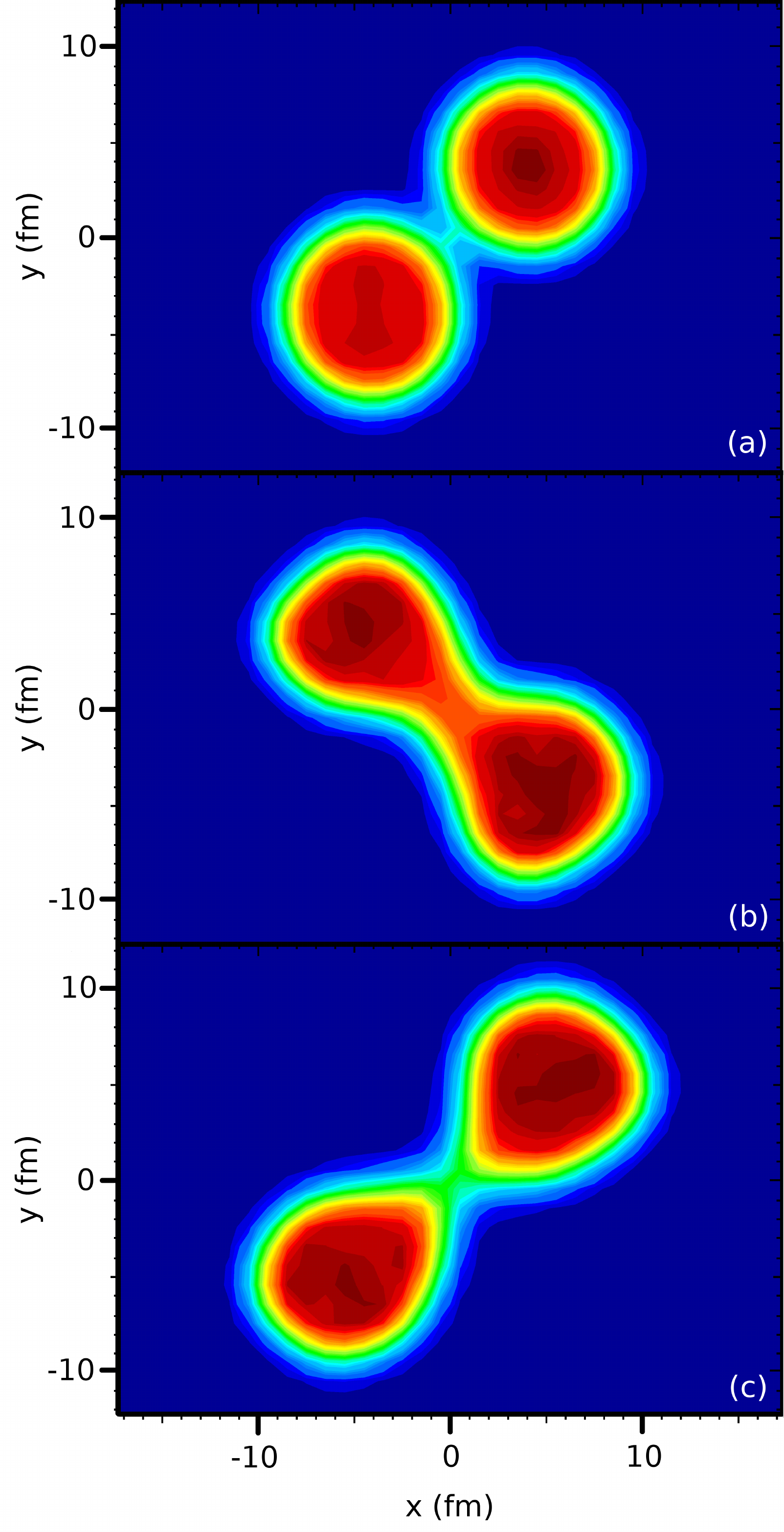}
\caption{(color online) Density profile of the ${}^{58} \text{Ni}+{}^{60} \text{Ni}$ at the center
of mass energy $E_\mathrm{c.m.} =135.6$~MeV and the impact parameter $b=5.2$~fm in the side
configuration, at times 300 (a), 750 (b) and 1250~fm/$c$ (c).}
\label{fig1}
\end{figure}
configurations of ${}^{58} \text{Ni}$ nucleus. As an example, Fig.~\ref{fig1} shows the density profile of
the ${}^{58} \text{Ni}+{}^{60} \text{Ni}$ at the center of mass energy of $E_\mathrm{c.m.} =135.6$~MeV
and the impact parameter $b=5.2$~fm for the initial side orientation of ${}^{58} \text{Ni}$, 
at times 300, 750 and 1250~fm/$c$. The dynamical symmetry axis of the di-nuclear
system is determined by the principle axis of the mass quadrupole moment tensor. The window plane is
perpendicular to the symmetry axis and passing through the minimum density slice. The neutron and
proton numbers on one side of the di-nuclear system (we refer to as projectile-like fragment) is
determined by integrating the local density on one side of the window plane. From the Langevin
equations of the neutron $N^{\lambda}$ and proton $Z^{\lambda}$ numbers of the projectile-like
fragments in each event, we can deduce a set of couple differential equations for the co-variances
$\sigma _{NN}^{2} (t)=\overline{(N^{\lambda } -N)^{2} }$, $\sigma _{ZZ}^{2}
(t)=\overline{(Z^{\lambda} -Z)^{2} }$, and $\sigma _{NZ}^{2} (t)=\overline{(N^{\lambda}
-N)(Z^{\lambda } -Z)}$. Here, $N=\overline{N^{\lambda}}$ and $Z=\overline{Z^{\lambda}}$ are mean values of the neutron and proton numbers, 
$\lambda$ indicates the event label and the bar denotes the average
over the ensemble generated in the SMF simulations. The coupled differential equations for the
co-variances are given by Eqs.~(17-19) in Ref.~\cite{ayik2018}, which involve the neutron $D_{NN} (t)$ and proton $D_{ZZ}
(t)$ diffusion coefficients and the derivatives of drift coefficients.
These set of coupled equations for covariances are also familiar from the
phenomenological nucleon exchange model, and they were
derived from the Fokker-Planck equation for the fragment
neutron and proton distributions in deep-inelastic heavy-ion 
collisions~\cite{schroder1981,merchant1981,merchant1982}.

\begin{figure}[!hpt]
\includegraphics*[width=8.5cm]{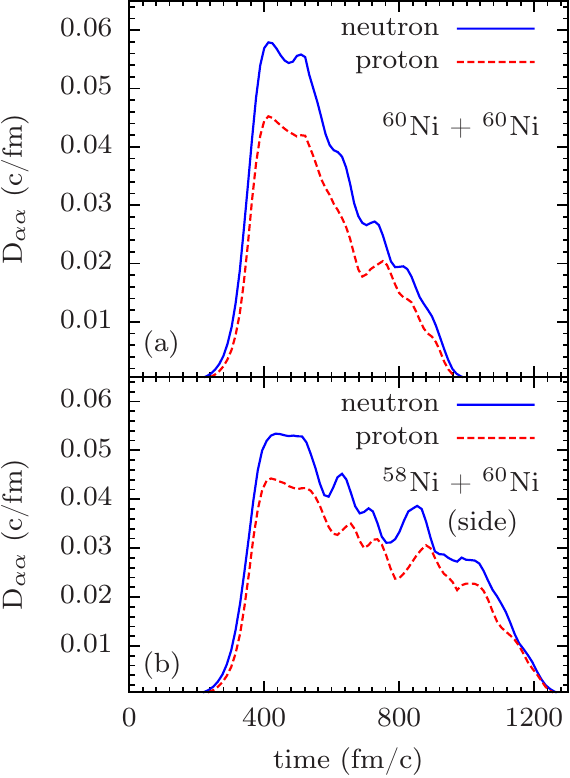}
\caption{(color online) Neutron (solid lines) and proton (dotted lines) diffusion coefficients at
the center of mass energy $E_\mathrm{c.m.} =135.6$~MeV and the impact parameter $b=5.2$~fm in
${}^{60} \text{Ni}+{}^{60} \text{Ni}$ collisions (a)  and in ${}^{58} \text{Ni}+{}^{60} \text{Ni}$
collisions in the side configuration (b).}
\label{fig2}
\end{figure}

\begin{figure}[!hpt]
\includegraphics*[width=8.5cm]{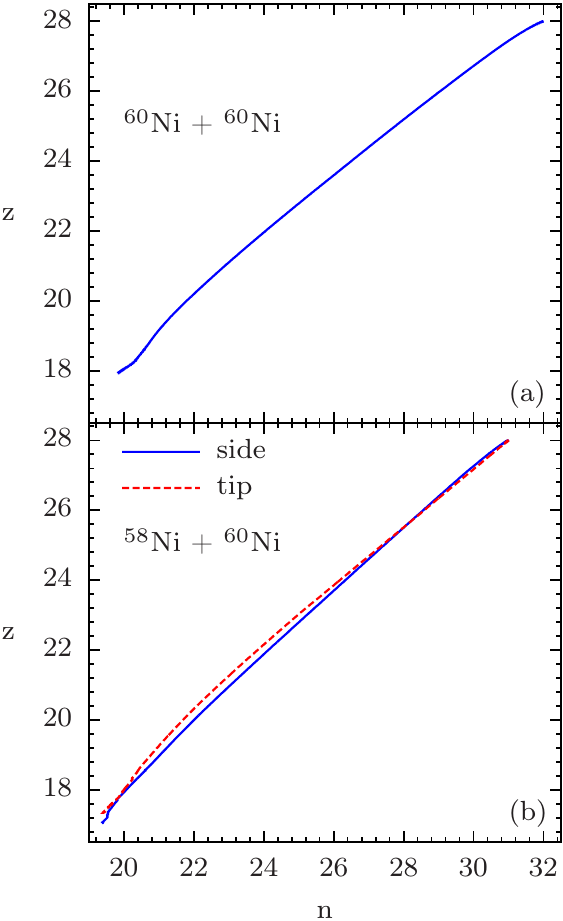}
\caption{(color online) One-sided mean-drift path at the center of mass energy $E_\mathrm{c.m.}
=135.6$~MeV and the impact parameter $b=5.2$~fm in ${}^{60} \text{Ni}+{}^{60} \text{Ni}$ collisions 
(a) and in ${}^{58} \text{Ni}+{}^{60} \text{Ni}$ collisions  in the side and the tip configurations 
(solid  and dashed lines) (b).}
\label{fig3}
\end{figure}

\begin{figure}[!hpt]
\includegraphics*[width=8.5cm]{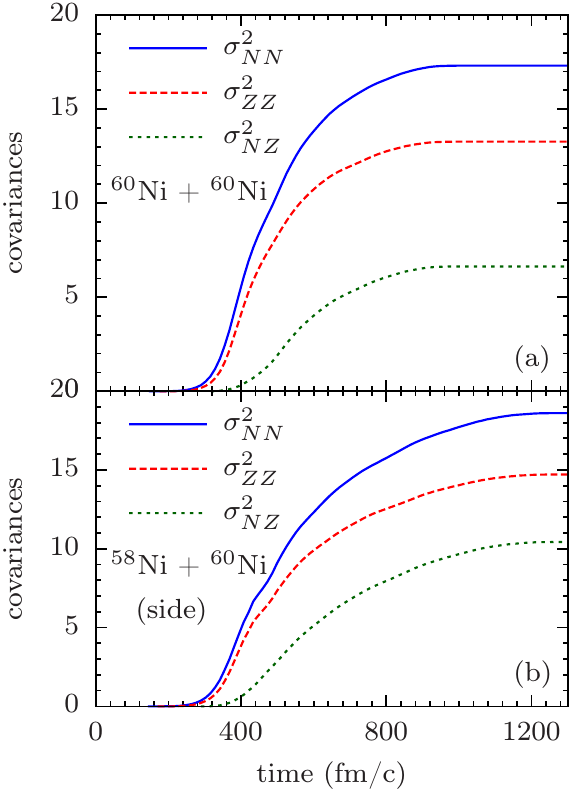}
\caption{(color online) Co-variances in the collision of ${}^{60} \text{Ni}+{}^{60} \text{Ni}$ (a)
and in the collision of ${}^{58} \text{Ni}+{}^{60} \text{Ni}$ in the side geometry (b) at the center
of mass  energy $E_\mathrm{c.m.} =135.6$~MeV and the impact parameter $b=5.2$~fm, or equivalently
initial orbital angular momentum $\ell=73\hbar$.}
\label{fig4}
\end{figure}

\begin{figure}[!hpt]
\includegraphics*[width=8.5cm]{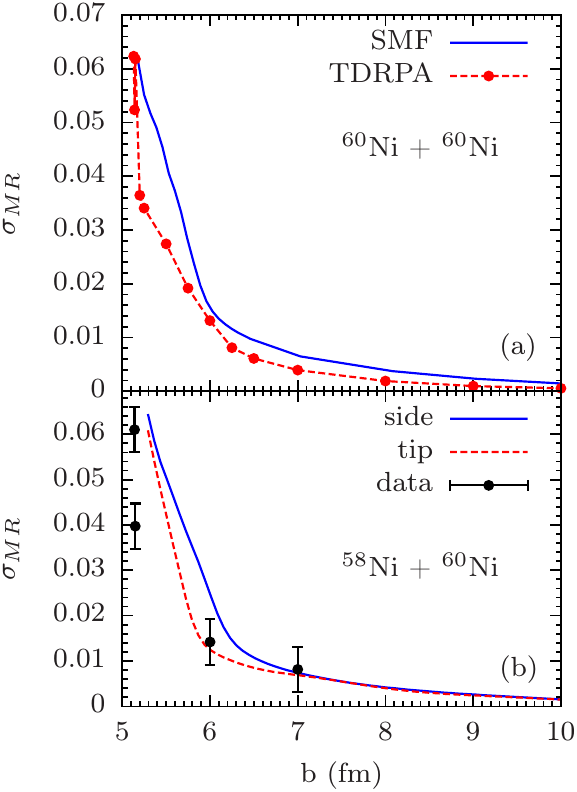}
\caption{(color online) Dispersion $\sigma _{MR}$ per unit mass as function of the impact parameter.
In (a), SMF and DRPA calculations in ${}^{60} \text{Ni}+{}^{60} \text{Ni}$ collisions are indicated
by solid line and dash line with dots, respectively. In (b) SMF calculations in side and tip
configurations (solid and dash lines) are compared with data.}
\label{fig5}
\end{figure}

For numerical calculations we employ the TDHF code of
Umar \textit{et al.}~\cite{umar1991a,umar2006c}. 
The computations are carried out in a box with size of $70\times 30\times 50$~fm$^3$. 
The initial separation of the nuclei is taken as 30~fm. The SLy4d~\cite{kim1997} Skyrme interaction is used.
We evaluate the diffusion coefficients using the Eq.~(37) in
Ref.~\cite{ayik2018} and determine the derivative of the drift coefficients around their mean values
from the one-sided drift path as described in Ref.~\cite{ayik2017}. Figure~\ref{fig2} shows neutron (solid
lines) and proton (dotted lines) diffusion coefficients at the center of mass energy
$E_\mathrm{c.m.} =135.6$~MeV and the impact parameter $b=5.2$~fm in ${}^{60} \text{Ni}+{}^{60}
\text{Ni}$ collisions in the upper panel and in ${}^{58} \text{Ni}+{}^{60} \text{Ni}$ collisions in
the side configuration in the lower panel.  
The fluctuations in the behavior of the diffusion coefficients are partly due to the effect of the
shell structure and partly due to the effect of the Pauli blocking of the occupied single particle
states. We note that, the contact time in the collision of ${}^{60} \text{Ni}+{}^{60} \text{Ni}$ is
about $600$~fm/$c$, which is shorter than the contact time of about $800$~fm/$c$ in the side collision
of the ${}^{58} \text{Ni}+{}^{60} \text{Ni}$ system. 
Figure~\ref{fig3} shows the one-sided mean drift path at the center of
mass energy $E_\mathrm{c.m.} =135.6$~MeV and the impact parameter $b=5.2$~fm in ${}^{60}
\text{Ni}+{}^{60} \text{Ni}$ collisions  (a) and in ${}^{58} \text{Ni}+{}^{60} \text{Ni}$
collisions  in the side configuration (solid line) and in the tip configuration (dashed line) (b). 
Here $n=N_{0} -N_{1}$ and   $z=Z_{0} -Z_{1} $.  The quantity $(N_{0} ,Z_{0})$ indicates the
equilibrium values of neutron and proton numbers which are $(32,28)$ for ${}^{60} \text{Ni}+{}^{60}
\text{Ni}$ and $(31,28)$ for ${}^{58} \text{Ni}+{}^{60} \text{Ni}$. The quantity $(N_{1} ,Z_{1})$
indicates the neutron and proton numbers of the fragment which are increasing due to gaining flux from its partner.  
Figure~\ref{fig4} shows the co-variances in the collision of ${}^{60} \text{Ni}+{}^{60} \text{Ni}$
in the upper panel (a) and in the collision of ${}^{58} \text{Ni}+{}^{60} \text{Ni}$ (in the side
geometry) in the lower panel (b) at the center of mass energy $E_\mathrm{c.m.} =135.6$~MeV and
impact parameter $b=5.2$~fm, or equivalently initial orbital angular momentum $\ell=73\hbar$.
Neutron-neutron $\sigma _{NN}^{2}$, proton-proton $\sigma _{ZZ}^{2}$, and neutron-proton $\sigma
_{NZ}^{2}$ co-variances are indicated by solid, dashed and dotted lines, respectively. Since for the
side geometry of the ${}^{58} \text{Ni}+{}^{60} \text{Ni}$ system, contact time is longer than the
spherical ${}^{60} \text{Ni}+{}^{60} \text{Ni}$ system, even though the center of mass energy and
the impact parameters are nearly the same, the co-variances are slightly larger in the ${}^{58}
\text{Ni}+{}^{60} \text{Ni}$ system. The mass number variance is determined as $\sigma _{AA}^{2}
=\sigma _{NN}^{2} +\sigma _{ZZ}^{2} +2\sigma _{NZ}^{2}$. 
In the upper panel of Fig.~\ref{fig5} (a), we compare the result of SMF calculations (solid line) for the mass
number dispersion per unit nucleon $\sigma _{MR} =\sigma _{AA} /A_{T} $, where $A_{T}$ is the total
mass number of the system,  with the calculations carried out in the TDRPA (dashed line with dots)
framework as a function of the impact parameters in the collision of the system ${}^{60}
\text{Ni}+{}^{60} \text{Ni}$ at the same center of mass energy $E_\mathrm{c.m.} =135.6$~MeV. Even
though the same Skyrme force, SLy4d, is used in both calculations, the SMF calculations give up to
30\% larger value than the TDRPA results for the dispersion in the impact parameter interval
$b=(5.2-5.6)$~fm.  The lower panel of Fig.~\ref{fig5} (b) shows a comparison of the SMF calculations
of $\sigma _{MR} =\sigma _{AA} /A_{T} $ for the side (solid line) and the tip (dashed line)
configurations with data as a function of the impact parameters. There are four data points that are
reported in Fig.~\ref{fig3} of Ref.~\cite{williams2018}. These points are indicated in the figure including experimental error bars. 
The SMF calculations with the side configurations
provide a better fit to the data. However, the data seems to indicate nearly two different magnitudes
for the mass dispersion at the impact parameter $b=5.2$~fm.   

\begin{figure}[!hpt]
\includegraphics*[width=8.5cm]{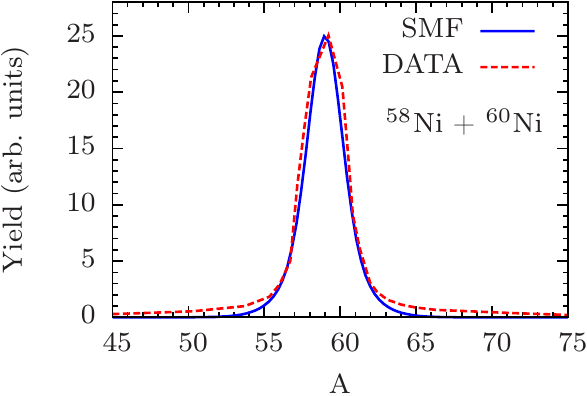}
\caption{(color online) Solid line is the yield obtained from the SMF calculations, which is
averaged over the tip and side configurations. The experimental yield is indicated by the dashed
line.}
\label{fig6}
\end{figure}

The stochastic Langevin dynamics of a set of macroscopic variables is equivalent to the
Fokker-Planck description of the distribution function of the macroscopic
variables~\cite{risken1996,gardiner1991,weiss1999}. When the driving potential energy has a
parabolic form, distribution function is a correlated Gaussian of macroscopic variables.  Here, the
macroscopic variable is the mass number $A$ of projectile-like fragment.  For a given impact parameter
$b$ or the initial orbital angular momentum $\ell$, the fragment mass distribution is given by a
Gaussian function, 
\begin{equation} \label{eq1} 
F_{\ell} (A)=\frac{1}{\sqrt{2\pi } } \frac{1}{\sigma _{AA} (\ell)} \exp \left[-\frac{1}{2}
\left(\frac{A-A(\ell)}{\sigma _{AA} (\ell)} \right)^{2} \right] \;,
\end{equation} 
where $A(\ell)$ and $\sigma _{AA}(\ell)$ denote the mean value and the dispersion of the mass number
in the collision with initial orbital angular momentum $\ell$.  In order to calculate the mass
distribution in the collision ${}^{58} \text{Ni}+{}^{60} \text{Ni}$ at center of mass energy
$E_\mathrm{c.m.} =135.6$~MeV, we average the Gaussian distribution given by Eq.~\eqref{eq1} over a
range of initial orbital angular momentum $\ell_{0} \le \ell\le \ell_{m} $, where $\ell_{0}
=73\hbar$ is the lowest initial orbital which does not lead to fusion and $\ell_{m} =96\hbar$ is
the maximum angular momentum corresponding to the detector resolution limit. The mass distribution
of the primary fragment is given by the weighted average of the Gaussian functions, 
\begin{equation} \label{eq2} 
P(A)=\frac{\eta }{\sum _{\ell}(2\ell+1) } \sum _{\ell}(2\ell+1)F_{\ell} (A)  \;,
\end{equation} 
where $\eta$ is a normalization constant. The solid line in Fig.~\ref{fig6} shows the yield obtained from the
SMF calculations which is averaged over the tip and side configurations. The experimental data is
indicated by the dashed line. Since the system is very close to symmetry, the mass asymmetry reaches
the equilibrium value in a very short time interval. In Eq.~\eqref{eq2} we take the equilibrium value
$A=59$ for each initial orbital angular momentum in the interval where the summation is carried
out.  We determined the normalization constant $\eta$ by matching the peak value of the experimental
yield at $A=59$ by matching the peak value of the experimental yield at $A=59$. The
experimental yield indicated by the dashed line, is deduced from the part (a) of the Fig.~1 in
Ref.~\cite{williams2018} with the data taken at the energy $E_\mathrm{c.m.}/V_B=1.4$. 

In conclusions, we find that the quantal diffusion description deduced from the SMF approach
provide a good description for the fragment mass distribution observed in ${}^{58} \text{Ni}+{}^{60}
\text{Ni}$ collisions at the center of mass energy $E_\mathrm{c.m.}=135.6$~MeV without any
adjustable parameters except the parameters of the effective Skyrme interaction in the TDHF code. 
Since in small impact parameter range, the mass dispersions take rather large values, the partition of 
the integrated data into small impact parameter range may introduce large experimental errors.  The
sizable discrepancy between the SMF calculations and the data from the part (b) of Fig.~\ref{fig5} may arise
from such partitioning of the integrated experimental data, in particular  in the small impact
parameter range.

S.A. gratefully acknowledges the IPN-Orsay and the Middle East Technical University for warm
hospitality extended to him during his visits. S.A. also gratefully acknowledges useful discussions
with D. Lacroix, and very much thankful to F. Ayik for continuous support and encouragement. This
work is supported in part by US DOE Grants Nos. DE-SC0015513 and
DE-SC0013847, and in part by TUBITAK Grant No. 117F109.

\bibliography{VU_bibtex_master}

\begin{thebibliography}{63}%
\makeatletter
\providecommand \@ifxundefined [1]{%
 \@ifx{#1\undefined}
}%
\providecommand \@ifnum [1]{%
 \ifnum #1\expandafter \@firstoftwo
 \else \expandafter \@secondoftwo
 \fi
}%
\providecommand \@ifx [1]{%
 \ifx #1\expandafter \@firstoftwo
 \else \expandafter \@secondoftwo
 \fi
}%
\providecommand \natexlab [1]{#1}%
\providecommand \enquote  [1]{``#1''}%
\providecommand \bibnamefont  [1]{#1}%
\providecommand \bibfnamefont [1]{#1}%
\providecommand \citenamefont [1]{#1}%
\providecommand \href@noop [0]{\@secondoftwo}%
\providecommand \href [0]{\begingroup \@sanitize@url \@href}%
\providecommand \@href[1]{\@@startlink{#1}\@@href}%
\providecommand \@@href[1]{\endgroup#1\@@endlink}%
\providecommand \@sanitize@url [0]{\catcode `\\12\catcode `\$12\catcode
  `\&12\catcode `\#12\catcode `\^12\catcode `\_12\catcode `\%12\relax}%
\providecommand \@@startlink[1]{}%
\providecommand \@@endlink[0]{}%
\providecommand \url  [0]{\begingroup\@sanitize@url \@url }%
\providecommand \@url [1]{\endgroup\@href {#1}{\urlprefix }}%
\providecommand \urlprefix  [0]{URL }%
\providecommand \Eprint [0]{\href }%
\providecommand \doibase [0]{http://dx.doi.org/}%
\providecommand \selectlanguage [0]{\@gobble}%
\providecommand \bibinfo  [0]{\@secondoftwo}%
\providecommand \bibfield  [0]{\@secondoftwo}%
\providecommand \translation [1]{[#1]}%
\providecommand \BibitemOpen [0]{}%
\providecommand \bibitemStop [0]{}%
\providecommand \bibitemNoStop [0]{.\EOS\space}%
\providecommand \EOS [0]{\spacefactor3000\relax}%
\providecommand \BibitemShut  [1]{\csname bibitem#1\endcsname}%
\let\auto@bib@innerbib\@empty
\bibitem [{\citenamefont {{Ph. Chomaz}}\ \emph {et~al.}(1993)\citenamefont
  {{Ph. Chomaz}}, \citenamefont {{Di Toro}},\ and\ \citenamefont
  {Smerzi}}]{chomaz1993}%
  \BibitemOpen
  \bibfield  {author} {\bibinfo {author} {\bibnamefont {{Ph. Chomaz}}},
  \bibinfo {author} {\bibfnamefont {M.}~\bibnamefont {{Di Toro}}}, \ and\
  \bibinfo {author} {\bibfnamefont {A.}~\bibnamefont {Smerzi}},\ }\bibfield
  {title} {\enquote {\bibinfo {title} {{P}re-equilibrium effects on properties
  of hot giant-dipole resonances},}\ }\href {\doibase
  10.1016/0375-9474(93)90126-I} {\bibfield  {journal} {\bibinfo  {journal}
  {Nucl. Phys. A}\ }\textbf {\bibinfo {volume} {563}},\ \bibinfo {pages}
  {509--524} (\bibinfo {year} {1993})}\BibitemShut {NoStop}%
\bibitem [{\citenamefont {Dasso}\ \emph {et~al.}(1995)\citenamefont {Dasso},
  \citenamefont {Pollarolo},\ and\ \citenamefont {Winther}}]{dasso1995}%
  \BibitemOpen
  \bibfield  {author} {\bibinfo {author} {\bibfnamefont {C.~H.}\ \bibnamefont
  {Dasso}}, \bibinfo {author} {\bibfnamefont {G.}~\bibnamefont {Pollarolo}}, \
  and\ \bibinfo {author} {\bibfnamefont {A.}~\bibnamefont {Winther}},\
  }\bibfield  {title} {\enquote {\bibinfo {title} {Particle evaporation
  following multinucleon transfer processes with radioactive beams},}\ }\href
  {\doibase 10.1103/physrevc.52.2264} {\bibfield  {journal} {\bibinfo
  {journal} {Phys. Rev. C}\ }\textbf {\bibinfo {volume} {52}},\ \bibinfo
  {pages} {2264--2265} (\bibinfo {year} {1995})}\BibitemShut {NoStop}%
\bibitem [{\citenamefont {Flibotte}\ \emph {et~al.}(1996)\citenamefont
  {Flibotte}, \citenamefont {Chomaz}, \citenamefont {Colonna}, \citenamefont
  {Cromaz}, \citenamefont {DeGraaf}, \citenamefont {Drake}, \citenamefont
  {Galindoi-Uribarri}, \citenamefont {Janzen}, \citenamefont {Jonkman},
  \citenamefont {Marshall}, \citenamefont {Mullins}, \citenamefont {Nieminen},
  \citenamefont {Radford}, \citenamefont {Rodriguez}, \citenamefont
  {Waddington}, \citenamefont {Ward},\ and\ \citenamefont
  {Wilson}}]{flibotte1996}%
  \BibitemOpen
  \bibfield  {author} {\bibinfo {author} {\bibfnamefont {S.}~\bibnamefont
  {Flibotte}}, \bibinfo {author} {\bibfnamefont {Ph.}\ \bibnamefont {Chomaz}},
  \bibinfo {author} {\bibfnamefont {M.}~\bibnamefont {Colonna}}, \bibinfo
  {author} {\bibfnamefont {M.}~\bibnamefont {Cromaz}}, \bibinfo {author}
  {\bibfnamefont {J.}~\bibnamefont {DeGraaf}}, \bibinfo {author} {\bibfnamefont
  {T.~E.}\ \bibnamefont {Drake}}, \bibinfo {author} {\bibfnamefont
  {A.}~\bibnamefont {Galindoi-Uribarri}}, \bibinfo {author} {\bibfnamefont
  {V.~P.}\ \bibnamefont {Janzen}}, \bibinfo {author} {\bibfnamefont
  {J.}~\bibnamefont {Jonkman}}, \bibinfo {author} {\bibfnamefont {S.~W.}\
  \bibnamefont {Marshall}}, \bibinfo {author} {\bibfnamefont {S.~M.}\
  \bibnamefont {Mullins}}, \bibinfo {author} {\bibfnamefont {J.~M.}\
  \bibnamefont {Nieminen}}, \bibinfo {author} {\bibfnamefont {D.~C.}\
  \bibnamefont {Radford}}, \bibinfo {author} {\bibfnamefont {J.~L.}\
  \bibnamefont {Rodriguez}}, \bibinfo {author} {\bibfnamefont {J.~C.}\
  \bibnamefont {Waddington}}, \bibinfo {author} {\bibfnamefont
  {D.}~\bibnamefont {Ward}}, \ and\ \bibinfo {author} {\bibfnamefont {J.~N.}\
  \bibnamefont {Wilson}},\ }\bibfield  {title} {\enquote {\bibinfo {title}
  {{P}re-equilibrium {E}ffects in the {P}opulation of {G}iant {D}ipole
  {R}esonances},}\ }\href {\doibase 10.1103/PhysRevLett.77.1448} {\bibfield
  {journal} {\bibinfo  {journal} {Phys. Rev. Lett.}\ }\textbf {\bibinfo
  {volume} {77}},\ \bibinfo {pages} {1448--1451} (\bibinfo {year}
  {1996})}\BibitemShut {NoStop}%
\bibitem [{\citenamefont {Baran}\ \emph {et~al.}(2001)\citenamefont {Baran},
  \citenamefont {Brink}, \citenamefont {Colonna},\ and\ \citenamefont {{Di
  Toro}}}]{baran2001}%
  \BibitemOpen
  \bibfield  {author} {\bibinfo {author} {\bibfnamefont {V.}~\bibnamefont
  {Baran}}, \bibinfo {author} {\bibfnamefont {D.~M.}\ \bibnamefont {Brink}},
  \bibinfo {author} {\bibfnamefont {M.}~\bibnamefont {Colonna}}, \ and\
  \bibinfo {author} {\bibfnamefont {M.}~\bibnamefont {{Di Toro}}},\ }\bibfield
  {title} {\enquote {\bibinfo {title} {{C}ollective {D}ipole {B}remsstrahlung
  in {F}usion {R}eactions},}\ }\href {\doibase 10.1103/PhysRevLett.87.182501}
  {\bibfield  {journal} {\bibinfo  {journal} {Phys. Rev. Lett.}\ }\textbf
  {\bibinfo {volume} {87}},\ \bibinfo {pages} {182501} (\bibinfo {year}
  {2001})}\BibitemShut {NoStop}%
\bibitem [{\citenamefont {Simenel}\ \emph {et~al.}(2001)\citenamefont
  {Simenel}, \citenamefont {Chomaz},\ and\ \citenamefont {{de
  France}}}]{simenel2001}%
  \BibitemOpen
  \bibfield  {author} {\bibinfo {author} {\bibfnamefont {C.}~\bibnamefont
  {Simenel}}, \bibinfo {author} {\bibfnamefont {Ph.}\ \bibnamefont {Chomaz}}, \
  and\ \bibinfo {author} {\bibfnamefont {G.}~\bibnamefont {{de France}}},\
  }\bibfield  {title} {\enquote {\bibinfo {title} {{Q}uantum {C}alculation of
  the {D}ipole {E}xcitation in {F}usion {R}eactions},}\ }\href {\doibase
  10.1103/PhysRevLett.86.2971} {\bibfield  {journal} {\bibinfo  {journal}
  {Phys. Rev. Lett.}\ }\textbf {\bibinfo {volume} {86}},\ \bibinfo {pages}
  {2971--2974} (\bibinfo {year} {2001})}\BibitemShut {NoStop}%
\bibitem [{\citenamefont {Back}\ \emph {et~al.}(2014)\citenamefont {Back},
  \citenamefont {Esbensen}, \citenamefont {Jiang},\ and\ \citenamefont
  {Rehm}}]{back2014}%
  \BibitemOpen
  \bibfield  {author} {\bibinfo {author} {\bibfnamefont {B.~B.}\ \bibnamefont
  {Back}}, \bibinfo {author} {\bibfnamefont {H.}~\bibnamefont {Esbensen}},
  \bibinfo {author} {\bibfnamefont {C.~L.}\ \bibnamefont {Jiang}}, \ and\
  \bibinfo {author} {\bibfnamefont {K.~E.}\ \bibnamefont {Rehm}},\ }\bibfield
  {title} {\enquote {\bibinfo {title} {{R}ecent developments in heavy-ion
  fusion reactions},}\ }\href {\doibase 10.1103/RevModPhys.86.317} {\bibfield
  {journal} {\bibinfo  {journal} {Rev. Mod. Phys.}\ }\textbf {\bibinfo {volume}
  {86}},\ \bibinfo {pages} {317--360} (\bibinfo {year} {2014})}\BibitemShut
  {NoStop}%
\bibitem [{\citenamefont {Broglia}\ \emph {et~al.}(1983)\citenamefont
  {Broglia}, \citenamefont {Dasso}, \citenamefont {Landowne},\ and\
  \citenamefont {Winther}}]{broglia1983}%
  \BibitemOpen
  \bibfield  {author} {\bibinfo {author} {\bibfnamefont {R.~A.}\ \bibnamefont
  {Broglia}}, \bibinfo {author} {\bibfnamefont {C.~H.}\ \bibnamefont {Dasso}},
  \bibinfo {author} {\bibfnamefont {S.}~\bibnamefont {Landowne}}, \ and\
  \bibinfo {author} {\bibfnamefont {A.}~\bibnamefont {Winther}},\ }\bibfield
  {title} {\enquote {\bibinfo {title} {{P}ossible effect of transfer reactions
  on heavy ion fusion at sub-barrier energies},}\ }\href {\doibase
  10.1103/physrevc.27.2433} {\bibfield  {journal} {\bibinfo  {journal} {Phys.
  Rev. C}\ }\textbf {\bibinfo {volume} {27}},\ \bibinfo {pages} {2433--2435}
  (\bibinfo {year} {1983})}\BibitemShut {NoStop}%
\bibitem [{\citenamefont {Esbensen}\ and\ \citenamefont
  {Landowne}(1989)}]{esbensen1989b}%
  \BibitemOpen
  \bibfield  {author} {\bibinfo {author} {\bibfnamefont {H.}~\bibnamefont
  {Esbensen}}\ and\ \bibinfo {author} {\bibfnamefont {S.}~\bibnamefont
  {Landowne}},\ }\bibfield  {title} {\enquote {\bibinfo {title}
  {{C}oupled-channels calculations for transfer reactions},}\ }\href {\doibase
  10.1016/0375-9474(89)90409-0} {\bibfield  {journal} {\bibinfo  {journal}
  {Nucl. Phys. A}\ }\textbf {\bibinfo {volume} {492}},\ \bibinfo {pages}
  {473--492} (\bibinfo {year} {1989})}\BibitemShut {NoStop}%
\bibitem [{\citenamefont {Morton}\ \emph {et~al.}(1994)\citenamefont {Morton},
  \citenamefont {Dasgupta}, \citenamefont {Hinde}, \citenamefont {Leigh},
  \citenamefont {Lemmon}, \citenamefont {Lestone}, \citenamefont {Mein},
  \citenamefont {Newton}, \citenamefont {Timmers}, \citenamefont {Rowley},\
  and\ \citenamefont {Kruppa}}]{morton1994}%
  \BibitemOpen
  \bibfield  {author} {\bibinfo {author} {\bibfnamefont {C.~R.}\ \bibnamefont
  {Morton}}, \bibinfo {author} {\bibfnamefont {M.}~\bibnamefont {Dasgupta}},
  \bibinfo {author} {\bibfnamefont {D.~J.}\ \bibnamefont {Hinde}}, \bibinfo
  {author} {\bibfnamefont {J.~R.}\ \bibnamefont {Leigh}}, \bibinfo {author}
  {\bibfnamefont {R.~C.}\ \bibnamefont {Lemmon}}, \bibinfo {author}
  {\bibfnamefont {J.~P.}\ \bibnamefont {Lestone}}, \bibinfo {author}
  {\bibfnamefont {J.~C.}\ \bibnamefont {Mein}}, \bibinfo {author}
  {\bibfnamefont {J.~O.}\ \bibnamefont {Newton}}, \bibinfo {author}
  {\bibfnamefont {H.}~\bibnamefont {Timmers}}, \bibinfo {author} {\bibfnamefont
  {N.}~\bibnamefont {Rowley}}, \ and\ \bibinfo {author} {\bibfnamefont {A.~T.}\
  \bibnamefont {Kruppa}},\ }\bibfield  {title} {\enquote {\bibinfo {title}
  {Clear signatures of specific inelastic and transfer channels in the
  distribution of fusion barriers},}\ }\href {\doibase
  10.1103/PhysRevLett.72.4074} {\bibfield  {journal} {\bibinfo  {journal}
  {Phys. Rev. Lett.}\ }\textbf {\bibinfo {volume} {72}},\ \bibinfo {pages}
  {4074--4077} (\bibinfo {year} {1994})}\BibitemShut {NoStop}%
\bibitem [{\citenamefont {Timmers}\ \emph {et~al.}(1998)\citenamefont
  {Timmers}, \citenamefont {Ackermann}, \citenamefont {Beghini}, \citenamefont
  {Corradi}, \citenamefont {He}, \citenamefont {Montagnoli}, \citenamefont
  {Scarlassara}, \citenamefont {Stefanini},\ and\ \citenamefont
  {Rowley}}]{timmers1998}%
  \BibitemOpen
  \bibfield  {author} {\bibinfo {author} {\bibfnamefont {H.}~\bibnamefont
  {Timmers}}, \bibinfo {author} {\bibfnamefont {D.}~\bibnamefont {Ackermann}},
  \bibinfo {author} {\bibfnamefont {S.}~\bibnamefont {Beghini}}, \bibinfo
  {author} {\bibfnamefont {L.}~\bibnamefont {Corradi}}, \bibinfo {author}
  {\bibfnamefont {J.H.}\ \bibnamefont {He}}, \bibinfo {author} {\bibfnamefont
  {G.}~\bibnamefont {Montagnoli}}, \bibinfo {author} {\bibfnamefont
  {F.}~\bibnamefont {Scarlassara}}, \bibinfo {author} {\bibfnamefont {A.M.}\
  \bibnamefont {Stefanini}}, \ and\ \bibinfo {author} {\bibfnamefont
  {N.}~\bibnamefont {Rowley}},\ }\bibfield  {title} {\enquote {\bibinfo {title}
  {{A} case study of collectivity, transfer and fusion enhancement},}\ }\href
  {\doibase 10.1016/s0375-9474(98)00121-3} {\bibfield  {journal} {\bibinfo
  {journal} {Nucl. Phys. A}\ }\textbf {\bibinfo {volume} {633}},\ \bibinfo
  {pages} {421--445} (\bibinfo {year} {1998})}\BibitemShut {NoStop}%
\bibitem [{\citenamefont {Stefanini}\ \emph {et~al.}(2007)\citenamefont
  {Stefanini}, \citenamefont {Behera}, \citenamefont {Beghini}, \citenamefont
  {Corradi}, \citenamefont {Fioretto}, \citenamefont {Gadea}, \citenamefont
  {Montagnoli}, \citenamefont {Rowley}, \citenamefont {Scarlassara},
  \citenamefont {Szilner},\ and\ \citenamefont {Trotta}}]{stefanini2007}%
  \BibitemOpen
  \bibfield  {author} {\bibinfo {author} {\bibfnamefont {A.~M.}\ \bibnamefont
  {Stefanini}}, \bibinfo {author} {\bibfnamefont {B.~R.}\ \bibnamefont
  {Behera}}, \bibinfo {author} {\bibfnamefont {S.}~\bibnamefont {Beghini}},
  \bibinfo {author} {\bibfnamefont {L.}~\bibnamefont {Corradi}}, \bibinfo
  {author} {\bibfnamefont {E.}~\bibnamefont {Fioretto}}, \bibinfo {author}
  {\bibfnamefont {A.}~\bibnamefont {Gadea}}, \bibinfo {author} {\bibfnamefont
  {G.}~\bibnamefont {Montagnoli}}, \bibinfo {author} {\bibfnamefont
  {N.}~\bibnamefont {Rowley}}, \bibinfo {author} {\bibfnamefont
  {F.}~\bibnamefont {Scarlassara}}, \bibinfo {author} {\bibfnamefont
  {S.}~\bibnamefont {Szilner}}, \ and\ \bibinfo {author} {\bibfnamefont
  {M.}~\bibnamefont {Trotta}},\ }\bibfield  {title} {\enquote {\bibinfo {title}
  {{S}ub-barrier fusion of $^{40}${C}a + $^{94}${Z}r : {I}nterplay of phonon
  and transfer couplings},}\ }\href {\doibase 10.1103/physrevc.76.014610}
  {\bibfield  {journal} {\bibinfo  {journal} {Phys. Rev. C}\ }\textbf {\bibinfo
  {volume} {76}},\ \bibinfo {pages} {014610} (\bibinfo {year}
  {2007})}\BibitemShut {NoStop}%
\bibitem [{\citenamefont {Zagrebaev}(2003)}]{zagrebaev2003}%
  \BibitemOpen
  \bibfield  {author} {\bibinfo {author} {\bibfnamefont {V.~I.}\ \bibnamefont
  {Zagrebaev}},\ }\bibfield  {title} {\enquote {\bibinfo {title} {{S}ub-barrier
  fusion enhancement due to neutron transfer},}\ }\href {\doibase
  10.1103/physrevc.67.061601} {\bibfield  {journal} {\bibinfo  {journal} {Phys.
  Rev. C}\ }\textbf {\bibinfo {volume} {67}},\ \bibinfo {pages} {061601}
  (\bibinfo {year} {2003})}\BibitemShut {NoStop}%
\bibitem [{\citenamefont {{Valery Zagrebaev}}\ and\ \citenamefont {{Walter
  Greiner}}(2007{\natexlab{a}})}]{zagrebaev2007b}%
  \BibitemOpen
  \bibfield  {author} {\bibinfo {author} {\bibnamefont {{Valery Zagrebaev}}}\
  and\ \bibinfo {author} {\bibnamefont {{Walter Greiner}}},\ }\bibfield
  {title} {\enquote {\bibinfo {title} {{L}ow-energy collisions of heavy nuclei:
  dynamics of sticking, mass transfer and fusion},}\ }\href {\doibase
  10.1088/0954-3899/34/1/001} {\bibfield  {journal} {\bibinfo  {journal} {J.
  Phys. G}\ }\textbf {\bibinfo {volume} {34}},\ \bibinfo {pages} {1} (\bibinfo
  {year} {2007}{\natexlab{a}})}\BibitemShut {NoStop}%
\bibitem [{\citenamefont {Umar}\ \emph {et~al.}(2008)\citenamefont {Umar},
  \citenamefont {Oberacker},\ and\ \citenamefont {Maruhn}}]{umar2008a}%
  \BibitemOpen
  \bibfield  {author} {\bibinfo {author} {\bibfnamefont {A.~S.}\ \bibnamefont
  {Umar}}, \bibinfo {author} {\bibfnamefont {V.~E.}\ \bibnamefont {Oberacker}},
  \ and\ \bibinfo {author} {\bibfnamefont {J.~A.}\ \bibnamefont {Maruhn}},\
  }\bibfield  {title} {\enquote {\bibinfo {title} {{N}eutron transfer dynamics
  and doorway to fusion in time-dependent {H}artree-{F}ock theory},}\ }\href
  {\doibase 10.1140/epja/i2008-10614-6} {\bibfield  {journal} {\bibinfo
  {journal} {Eur. Phys. J. A}\ }\textbf {\bibinfo {volume} {37}},\ \bibinfo
  {pages} {245--250} (\bibinfo {year} {2008})}\BibitemShut {NoStop}%
\bibitem [{\citenamefont {Corradi}\ \emph {et~al.}(2009)\citenamefont
  {Corradi}, \citenamefont {Pollarolo},\ and\ \citenamefont
  {Szilner}}]{corradi2009}%
  \BibitemOpen
  \bibfield  {author} {\bibinfo {author} {\bibfnamefont {L.}~\bibnamefont
  {Corradi}}, \bibinfo {author} {\bibfnamefont {G.}~\bibnamefont {Pollarolo}},
  \ and\ \bibinfo {author} {\bibfnamefont {S.}~\bibnamefont {Szilner}},\
  }\bibfield  {title} {\enquote {\bibinfo {title} {Multinucleon transfer
  processes in heavy-ion reactions},}\ }\href {\doibase
  10.1088/0954-3899/36/11/113101} {\bibfield  {journal} {\bibinfo  {journal}
  {J. Phys. G}\ }\textbf {\bibinfo {volume} {36}},\ \bibinfo {pages} {113101}
  (\bibinfo {year} {2009})}\BibitemShut {NoStop}%
\bibitem [{\citenamefont {Zhang}\ \emph {et~al.}(2010)\citenamefont {Zhang},
  \citenamefont {Lin}, \citenamefont {Yang}, \citenamefont {Jia}, \citenamefont
  {Xu}, \citenamefont {Wu}, \citenamefont {Jia}, \citenamefont {Zhang},
  \citenamefont {Liu}, \citenamefont {Richard},\ and\ \citenamefont
  {Beck}}]{zhang2010}%
  \BibitemOpen
  \bibfield  {author} {\bibinfo {author} {\bibfnamefont {H.~Q.}\ \bibnamefont
  {Zhang}}, \bibinfo {author} {\bibfnamefont {C.~J.}\ \bibnamefont {Lin}},
  \bibinfo {author} {\bibfnamefont {F.}~\bibnamefont {Yang}}, \bibinfo {author}
  {\bibfnamefont {H.~M.}\ \bibnamefont {Jia}}, \bibinfo {author} {\bibfnamefont
  {X.~X.}\ \bibnamefont {Xu}}, \bibinfo {author} {\bibfnamefont {Z.~D.}\
  \bibnamefont {Wu}}, \bibinfo {author} {\bibfnamefont {F.}~\bibnamefont
  {Jia}}, \bibinfo {author} {\bibfnamefont {S.~T.}\ \bibnamefont {Zhang}},
  \bibinfo {author} {\bibfnamefont {Z.~H.}\ \bibnamefont {Liu}}, \bibinfo
  {author} {\bibfnamefont {A.}~\bibnamefont {Richard}}, \ and\ \bibinfo
  {author} {\bibfnamefont {C.}~\bibnamefont {Beck}},\ }\bibfield  {title}
  {\enquote {\bibinfo {title} {{N}ear-barrier fusion of $^{32}${S} +
  $^{90,96}${Z}r : {T}he effect of multi-neutron transfers in sub-barrier
  fusion reactions},}\ }\href {\doibase 10.1103/physrevc.82.054609} {\bibfield
  {journal} {\bibinfo  {journal} {Phys. Rev. C}\ }\textbf {\bibinfo {volume}
  {82}},\ \bibinfo {pages} {054609} (\bibinfo {year} {2010})}\BibitemShut
  {NoStop}%
\bibitem [{\citenamefont {Kohley}\ \emph {et~al.}(2011)\citenamefont {Kohley},
  \citenamefont {Liang}, \citenamefont {Shapira}, \citenamefont {Varner},
  \citenamefont {Gross}, \citenamefont {Allmond}, \citenamefont {Caraley},
  \citenamefont {Coello}, \citenamefont {Favela}, \citenamefont {Lagergren},\
  and\ \citenamefont {Mueller}}]{kohley2011}%
  \BibitemOpen
  \bibfield  {author} {\bibinfo {author} {\bibfnamefont {Z.}~\bibnamefont
  {Kohley}}, \bibinfo {author} {\bibfnamefont {J.~F.}\ \bibnamefont {Liang}},
  \bibinfo {author} {\bibfnamefont {D.}~\bibnamefont {Shapira}}, \bibinfo
  {author} {\bibfnamefont {R.~L.}\ \bibnamefont {Varner}}, \bibinfo {author}
  {\bibfnamefont {C.~J.}\ \bibnamefont {Gross}}, \bibinfo {author}
  {\bibfnamefont {J.~M.}\ \bibnamefont {Allmond}}, \bibinfo {author}
  {\bibfnamefont {A.~L.}\ \bibnamefont {Caraley}}, \bibinfo {author}
  {\bibfnamefont {E.~A.}\ \bibnamefont {Coello}}, \bibinfo {author}
  {\bibfnamefont {F.}~\bibnamefont {Favela}}, \bibinfo {author} {\bibfnamefont
  {K.}~\bibnamefont {Lagergren}}, \ and\ \bibinfo {author} {\bibfnamefont
  {P.~E.}\ \bibnamefont {Mueller}},\ }\bibfield  {title} {\enquote {\bibinfo
  {title} {{N}ear-{B}arrier {F}usion of $\mathrm{Sn}+\mathrm{Ni}$ and
  $\mathrm{Te}+\mathrm{Ni}$ {S}ystems: {E}xamining the {C}orrelation between
  {N}ucleon {T}ransfer and {F}usion {E}nhancement},}\ }\href {\doibase
  10.1103/PhysRevLett.107.202701} {\bibfield  {journal} {\bibinfo  {journal}
  {Phys. Rev. Lett.}\ }\textbf {\bibinfo {volume} {107}},\ \bibinfo {pages}
  {202701} (\bibinfo {year} {2011})}\BibitemShut {NoStop}%
\bibitem [{\citenamefont {Montagnoli}\ \emph {et~al.}(2013)\citenamefont
  {Montagnoli}, \citenamefont {Stefanini}, \citenamefont {Esbensen},
  \citenamefont {Jiang}, \citenamefont {Corradi}, \citenamefont {Courtin},
  \citenamefont {Fioretto}, \citenamefont {Goasduff}, \citenamefont {Grebosz},
  \citenamefont {Haas}, \citenamefont {Mazzocco}, \citenamefont {Michelagnoli},
  \citenamefont {Mijatovic}, \citenamefont {Montanari}, \citenamefont
  {Parascandolo}, \citenamefont {Rehm}, \citenamefont {Scarlassara},
  \citenamefont {Szilner}, \citenamefont {Tang},\ and\ \citenamefont
  {Ur}}]{montagnoli2013}%
  \BibitemOpen
  \bibfield  {author} {\bibinfo {author} {\bibfnamefont {G.}~\bibnamefont
  {Montagnoli}}, \bibinfo {author} {\bibfnamefont {A.~M.}\ \bibnamefont
  {Stefanini}}, \bibinfo {author} {\bibfnamefont {H.}~\bibnamefont {Esbensen}},
  \bibinfo {author} {\bibfnamefont {C.~L.}\ \bibnamefont {Jiang}}, \bibinfo
  {author} {\bibfnamefont {L.}~\bibnamefont {Corradi}}, \bibinfo {author}
  {\bibfnamefont {S.}~\bibnamefont {Courtin}}, \bibinfo {author} {\bibfnamefont
  {E.}~\bibnamefont {Fioretto}}, \bibinfo {author} {\bibfnamefont
  {A.}~\bibnamefont {Goasduff}}, \bibinfo {author} {\bibfnamefont
  {J.}~\bibnamefont {Grebosz}}, \bibinfo {author} {\bibfnamefont
  {F.}~\bibnamefont {Haas}}, \bibinfo {author} {\bibfnamefont {M.}~\bibnamefont
  {Mazzocco}}, \bibinfo {author} {\bibfnamefont {C.}~\bibnamefont
  {Michelagnoli}}, \bibinfo {author} {\bibfnamefont {T.}~\bibnamefont
  {Mijatovic}}, \bibinfo {author} {\bibfnamefont {D.}~\bibnamefont
  {Montanari}}, \bibinfo {author} {\bibfnamefont {C.}~\bibnamefont
  {Parascandolo}}, \bibinfo {author} {\bibfnamefont {K.~E.}\ \bibnamefont
  {Rehm}}, \bibinfo {author} {\bibfnamefont {F.}~\bibnamefont {Scarlassara}},
  \bibinfo {author} {\bibfnamefont {S.}~\bibnamefont {Szilner}}, \bibinfo
  {author} {\bibfnamefont {X.~D.}\ \bibnamefont {Tang}}, \ and\ \bibinfo
  {author} {\bibfnamefont {C.~A.}\ \bibnamefont {Ur}},\ }\bibfield  {title}
  {\enquote {\bibinfo {title} {Effects of transfer channels on near- and
  sub-barrier fusion of ${}^{\text{32}}${S} $+$ ${}^{\text{48}}${C}a},}\ }\href
  {\doibase 10.1103/PhysRevC.87.014611} {\bibfield  {journal} {\bibinfo
  {journal} {Phys. Rev. C}\ }\textbf {\bibinfo {volume} {87}},\ \bibinfo
  {pages} {014611} (\bibinfo {year} {2013})}\BibitemShut {NoStop}%
\bibitem [{\citenamefont {Gautam}(2014)}]{gautam2014}%
  \BibitemOpen
  \bibfield  {author} {\bibinfo {author} {\bibfnamefont {Manjeet~Singh}\
  \bibnamefont {Gautam}},\ }\bibfield  {title} {\enquote {\bibinfo {title}
  {Role of neutron transfer in the enhancement of sub-barrier fusion excitation
  functions of various systems using an energy-dependent {W}oods-{S}axon
  potential},}\ }\href {\doibase 10.1103/physrevc.90.024620} {\bibfield
  {journal} {\bibinfo  {journal} {Phys. Rev. C}\ }\textbf {\bibinfo {volume}
  {90}},\ \bibinfo {pages} {024620} (\bibinfo {year} {2014})}\BibitemShut
  {NoStop}%
\bibitem [{\citenamefont {Bourgin}\ \emph {et~al.}(2014)\citenamefont
  {Bourgin}, \citenamefont {Courtin}, \citenamefont {Haas}, \citenamefont
  {Stefanini}, \citenamefont {Montagnoli}, \citenamefont {Goasduff},
  \citenamefont {Montanari}, \citenamefont {Corradi}, \citenamefont {Fioretto},
  \citenamefont {Huiming}, \citenamefont {Scarlassara}, \citenamefont {Rowley},
  \citenamefont {Szilner},\ and\ \citenamefont
  {Mijatovi{\'{c}}}}]{bourgin2014}%
  \BibitemOpen
  \bibfield  {author} {\bibinfo {author} {\bibfnamefont {D.}~\bibnamefont
  {Bourgin}}, \bibinfo {author} {\bibfnamefont {S.}~\bibnamefont {Courtin}},
  \bibinfo {author} {\bibfnamefont {F.}~\bibnamefont {Haas}}, \bibinfo {author}
  {\bibfnamefont {A.~M.}\ \bibnamefont {Stefanini}}, \bibinfo {author}
  {\bibfnamefont {G.}~\bibnamefont {Montagnoli}}, \bibinfo {author}
  {\bibfnamefont {A.}~\bibnamefont {Goasduff}}, \bibinfo {author}
  {\bibfnamefont {D.}~\bibnamefont {Montanari}}, \bibinfo {author}
  {\bibfnamefont {L.}~\bibnamefont {Corradi}}, \bibinfo {author} {\bibfnamefont
  {E.}~\bibnamefont {Fioretto}}, \bibinfo {author} {\bibfnamefont
  {J.}~\bibnamefont {Huiming}}, \bibinfo {author} {\bibfnamefont
  {F.}~\bibnamefont {Scarlassara}}, \bibinfo {author} {\bibfnamefont
  {N.}~\bibnamefont {Rowley}}, \bibinfo {author} {\bibfnamefont
  {S.}~\bibnamefont {Szilner}}, \ and\ \bibinfo {author} {\bibfnamefont
  {T.}~\bibnamefont {Mijatovi{\'{c}}}},\ }\bibfield  {title} {\enquote
  {\bibinfo {title} {Barrier distributions and signatures of transfer channels
  in the $^{40}${C}a+$^{58,64}${N}i fusion reactions at energies around and
  below the {C}oulomb barrier},}\ }\href {\doibase 10.1103/physrevc.90.044610}
  {\bibfield  {journal} {\bibinfo  {journal} {Phys. Rev. C}\ }\textbf {\bibinfo
  {volume} {90}},\ \bibinfo {pages} {044610} (\bibinfo {year}
  {2014})}\BibitemShut {NoStop}%
\bibitem [{\citenamefont {Liang}\ \emph {et~al.}(2016)\citenamefont {Liang},
  \citenamefont {Allmond}, \citenamefont {Gross}, \citenamefont {Mueller},
  \citenamefont {Shapira}, \citenamefont {Varner}, \citenamefont {Dasgupta},
  \citenamefont {Hinde}, \citenamefont {Simenel}, \citenamefont {Williams},
  \citenamefont {{Vo--Phuoc}}, \citenamefont {Brown}, \citenamefont {Carter},
  \citenamefont {Evers}, \citenamefont {Luong}, \citenamefont {Ebadi},\ and\
  \citenamefont {Wakhle}}]{liang2016}%
  \BibitemOpen
  \bibfield  {author} {\bibinfo {author} {\bibfnamefont {J.~F.}\ \bibnamefont
  {Liang}}, \bibinfo {author} {\bibfnamefont {J.~M.}\ \bibnamefont {Allmond}},
  \bibinfo {author} {\bibfnamefont {C.~J.}\ \bibnamefont {Gross}}, \bibinfo
  {author} {\bibfnamefont {P.~E.}\ \bibnamefont {Mueller}}, \bibinfo {author}
  {\bibfnamefont {D.}~\bibnamefont {Shapira}}, \bibinfo {author} {\bibfnamefont
  {R.~L.}\ \bibnamefont {Varner}}, \bibinfo {author} {\bibfnamefont
  {M.}~\bibnamefont {Dasgupta}}, \bibinfo {author} {\bibfnamefont {D.~J.}\
  \bibnamefont {Hinde}}, \bibinfo {author} {\bibfnamefont {C.}~\bibnamefont
  {Simenel}}, \bibinfo {author} {\bibfnamefont {E.}~\bibnamefont {Williams}},
  \bibinfo {author} {\bibfnamefont {K.}~\bibnamefont {{Vo--Phuoc}}}, \bibinfo
  {author} {\bibfnamefont {M.~L.}\ \bibnamefont {Brown}}, \bibinfo {author}
  {\bibfnamefont {I.~P.}\ \bibnamefont {Carter}}, \bibinfo {author}
  {\bibfnamefont {M.}~\bibnamefont {Evers}}, \bibinfo {author} {\bibfnamefont
  {D.~H.}\ \bibnamefont {Luong}}, \bibinfo {author} {\bibfnamefont
  {T.}~\bibnamefont {Ebadi}}, \ and\ \bibinfo {author} {\bibfnamefont
  {A.}~\bibnamefont {Wakhle}},\ }\bibfield  {title} {\enquote {\bibinfo {title}
  {Examining the role of transfer coupling in sub-barrier fusion of
  $^{46,50}${T}i+$^{124}${S}n},}\ }\href {\doibase 10.1103/physrevc.94.024616}
  {\bibfield  {journal} {\bibinfo  {journal} {Phys. Rev. C}\ }\textbf {\bibinfo
  {volume} {94}},\ \bibinfo {pages} {024616} (\bibinfo {year}
  {2016})}\BibitemShut {NoStop}%
\bibitem [{\citenamefont {Volkov}(1978)}]{volkov1978}%
  \BibitemOpen
  \bibfield  {author} {\bibinfo {author} {\bibfnamefont {V.~V.}\ \bibnamefont
  {Volkov}},\ }\bibfield  {title} {\enquote {\bibinfo {title} {Deep inelastic
  transfer reactions -- {T}he new type of reactions between complex nuclei},}\
  }\href {\doibase 10.1016/0370-1573(78)90200-4} {\bibfield  {journal}
  {\bibinfo  {journal} {Phys. Rep.}\ }\textbf {\bibinfo {volume} {44}},\
  \bibinfo {pages} {93} (\bibinfo {year} {1978})}\BibitemShut {NoStop}%
\bibitem [{\citenamefont {Antonenko}\ and\ \citenamefont
  {Jolos}(1991)}]{antonenko1991}%
  \BibitemOpen
  \bibfield  {author} {\bibinfo {author} {\bibfnamefont {N.~V.}\ \bibnamefont
  {Antonenko}}\ and\ \bibinfo {author} {\bibfnamefont {R.~V.}\ \bibnamefont
  {Jolos}},\ }\bibfield  {title} {\enquote {\bibinfo {title} {The microscopic
  treatment of proton and neutron multiple transfer in {DIC}},}\ }\href
  {\doibase 10.1007/bf01295770} {\bibfield  {journal} {\bibinfo  {journal} {Z.
  Phys. A}\ }\textbf {\bibinfo {volume} {338}},\ \bibinfo {pages} {423--430}
  (\bibinfo {year} {1991})}\BibitemShut {NoStop}%
\bibitem [{\citenamefont {Adamian}\ \emph {et~al.}(2003)\citenamefont
  {Adamian}, \citenamefont {Antonenko},\ and\ \citenamefont
  {Scheid}}]{adamian2003}%
  \BibitemOpen
  \bibfield  {author} {\bibinfo {author} {\bibfnamefont {G.~G.}\ \bibnamefont
  {Adamian}}, \bibinfo {author} {\bibfnamefont {N.~V.}\ \bibnamefont
  {Antonenko}}, \ and\ \bibinfo {author} {\bibfnamefont {W.}~\bibnamefont
  {Scheid}},\ }\bibfield  {title} {\enquote {\bibinfo {title}
  {{C}haracteristics of quasifission products within the dinuclear system
  model},}\ }\href {\doibase 10.1103/PhysRevC.68.034601} {\bibfield  {journal}
  {\bibinfo  {journal} {Phys. Rev. C}\ }\textbf {\bibinfo {volume} {68}},\
  \bibinfo {pages} {034601} (\bibinfo {year} {2003})}\BibitemShut {NoStop}%
\bibitem [{\citenamefont {Aritomo}(2009)}]{aritomo2009}%
  \BibitemOpen
  \bibfield  {author} {\bibinfo {author} {\bibfnamefont {Y.}~\bibnamefont
  {Aritomo}},\ }\bibfield  {title} {\enquote {\bibinfo {title} {{A}nalysis of
  dynamical processes using the mass distribution of fission fragments in
  heavy-ion reactions},}\ }\href {\doibase 10.1103/PhysRevC.80.064604}
  {\bibfield  {journal} {\bibinfo  {journal} {Phys. Rev. C}\ }\textbf {\bibinfo
  {volume} {80}},\ \bibinfo {pages} {064604} (\bibinfo {year}
  {2009})}\BibitemShut {NoStop}%
\bibitem [{\citenamefont {Umar}\ \emph {et~al.}(2016)\citenamefont {Umar},
  \citenamefont {Oberacker},\ and\ \citenamefont {Simenel}}]{umar2016}%
  \BibitemOpen
  \bibfield  {author} {\bibinfo {author} {\bibfnamefont {A.~S.}\ \bibnamefont
  {Umar}}, \bibinfo {author} {\bibfnamefont {V.~E.}\ \bibnamefont {Oberacker}},
  \ and\ \bibinfo {author} {\bibfnamefont {C.}~\bibnamefont {Simenel}},\
  }\bibfield  {title} {\enquote {\bibinfo {title} {Fusion and quasifission
  dynamics in the reactions $^{48}\mathrm{Ca}+^{249}\mathrm{Bk}$ and
  $^{50}\mathrm{Ti}+^{249}\mathrm{Bk}$ using a time-dependent {H}artree-{F}ock
  approach},}\ }\href {\doibase 10.1103/PhysRevC.94.024605} {\bibfield
  {journal} {\bibinfo  {journal} {Phys. Rev. C}\ }\textbf {\bibinfo {volume}
  {94}},\ \bibinfo {pages} {024605} (\bibinfo {year} {2016})}\BibitemShut
  {NoStop}%
\bibitem [{\citenamefont {Evers}\ \emph {et~al.}(2011)\citenamefont {Evers},
  \citenamefont {Dasgupta}, \citenamefont {Hinde}, \citenamefont {Luong},
  \citenamefont {Rafiei}, \citenamefont {du~Rietz},\ and\ \citenamefont
  {Simenel}}]{evers2011}%
  \BibitemOpen
  \bibfield  {author} {\bibinfo {author} {\bibfnamefont {M.}~\bibnamefont
  {Evers}}, \bibinfo {author} {\bibfnamefont {M.}~\bibnamefont {Dasgupta}},
  \bibinfo {author} {\bibfnamefont {D.~J.}\ \bibnamefont {Hinde}}, \bibinfo
  {author} {\bibfnamefont {D.~H.}\ \bibnamefont {Luong}}, \bibinfo {author}
  {\bibfnamefont {R.}~\bibnamefont {Rafiei}}, \bibinfo {author} {\bibfnamefont
  {R.}~\bibnamefont {du~Rietz}}, \ and\ \bibinfo {author} {\bibfnamefont
  {C.}~\bibnamefont {Simenel}},\ }\bibfield  {title} {\enquote {\bibinfo
  {title} {{C}luster transfer in the reaction $^{16}${O}+$^{208}${P}b at
  energies well below the fusion barrier: {A} possible doorway to energy
  dissipation},}\ }\href {\doibase 10.1103/PhysRevC.84.054614} {\bibfield
  {journal} {\bibinfo  {journal} {Phys. Rev. C}\ }\textbf {\bibinfo {volume}
  {84}},\ \bibinfo {pages} {054614} (\bibinfo {year} {2011})}\BibitemShut
  {NoStop}%
\bibitem [{\citenamefont {{Washiyama}}\ and\ \citenamefont
  {{Lacroix}}(2009)}]{washiyama2009}%
  \BibitemOpen
  \bibfield  {author} {\bibinfo {author} {\bibfnamefont {K.}~\bibnamefont
  {{Washiyama}}}\ and\ \bibinfo {author} {\bibfnamefont {D.}~\bibnamefont
  {{Lacroix}}},\ }\bibfield  {title} {\enquote {\bibinfo {title} {Energy
  dissipation in fusion reactions from dynamical mean-field theory},}\ }\href
  {\doibase 10.1142/S0218301309014391} {\bibfield  {journal} {\bibinfo
  {journal} {Int. J. Mod. Phys. E}\ }\textbf {\bibinfo {volume} {18}},\
  \bibinfo {pages} {2114} (\bibinfo {year} {2009})}\BibitemShut {NoStop}%
\bibitem [{\citenamefont {Oberacker}\ \emph {et~al.}(2014)\citenamefont
  {Oberacker}, \citenamefont {Umar},\ and\ \citenamefont
  {Simenel}}]{oberacker2014}%
  \BibitemOpen
  \bibfield  {author} {\bibinfo {author} {\bibfnamefont {V.~E.}\ \bibnamefont
  {Oberacker}}, \bibinfo {author} {\bibfnamefont {A.~S.}\ \bibnamefont {Umar}},
  \ and\ \bibinfo {author} {\bibfnamefont {C.}~\bibnamefont {Simenel}},\
  }\bibfield  {title} {\enquote {\bibinfo {title} {{D}issipative dynamics in
  quasifission},}\ }\href {\doibase 10.1103/PhysRevC.90.054605} {\bibfield
  {journal} {\bibinfo  {journal} {Phys. Rev. C}\ }\textbf {\bibinfo {volume}
  {90}},\ \bibinfo {pages} {054605} (\bibinfo {year} {2014})}\BibitemShut
  {NoStop}%
\bibitem [{\citenamefont {Umar}\ \emph {et~al.}(2017)\citenamefont {Umar},
  \citenamefont {Simenel},\ and\ \citenamefont {Ye}}]{umar2017}%
  \BibitemOpen
  \bibfield  {author} {\bibinfo {author} {\bibfnamefont {A.~S.}\ \bibnamefont
  {Umar}}, \bibinfo {author} {\bibfnamefont {C.}~\bibnamefont {Simenel}}, \
  and\ \bibinfo {author} {\bibfnamefont {W.}~\bibnamefont {Ye}},\ }\bibfield
  {title} {\enquote {\bibinfo {title} {Transport properties of isospin
  asymmetric nuclear matter using the time-dependent {H}artree--{F}ock
  method},}\ }\href {\doibase 10.1103/PhysRevC.96.024625} {\bibfield  {journal}
  {\bibinfo  {journal} {Phys. Rev. C}\ }\textbf {\bibinfo {volume} {96}},\
  \bibinfo {pages} {024625} (\bibinfo {year} {2017})}\BibitemShut {NoStop}%
\bibitem [{\citenamefont {Wen}\ \emph {et~al.}(2018)\citenamefont {Wen},
  \citenamefont {Barton}, \citenamefont {Rios},\ and\ \citenamefont
  {Stevenson}}]{wen2018}%
  \BibitemOpen
  \bibfield  {author} {\bibinfo {author} {\bibfnamefont {Kai}\ \bibnamefont
  {Wen}}, \bibinfo {author} {\bibfnamefont {M.~C.}\ \bibnamefont {Barton}},
  \bibinfo {author} {\bibfnamefont {Arnau}\ \bibnamefont {Rios}}, \ and\
  \bibinfo {author} {\bibfnamefont {P.~D.}\ \bibnamefont {Stevenson}},\
  }\bibfield  {title} {\enquote {\bibinfo {title} {Two--body dissipation effect
  in nuclear fusion reactions},}\ }\href {\doibase 10.1103/PhysRevC.98.014603}
  {\bibfield  {journal} {\bibinfo  {journal} {Phys. Rev. C}\ }\textbf {\bibinfo
  {volume} {98}},\ \bibinfo {pages} {014603} (\bibinfo {year}
  {2018})}\BibitemShut {NoStop}%
\bibitem [{\citenamefont {Adamian}\ \emph {et~al.}(1996)\citenamefont
  {Adamian}, \citenamefont {Jolos}, \citenamefont {Nasirov},\ and\
  \citenamefont {Muminov}}]{adamian1996}%
  \BibitemOpen
  \bibfield  {author} {\bibinfo {author} {\bibfnamefont {G.~G.}\ \bibnamefont
  {Adamian}}, \bibinfo {author} {\bibfnamefont {R.~V.}\ \bibnamefont {Jolos}},
  \bibinfo {author} {\bibfnamefont {A.~K.}\ \bibnamefont {Nasirov}}, \ and\
  \bibinfo {author} {\bibfnamefont {A.~I.}\ \bibnamefont {Muminov}},\
  }\bibfield  {title} {\enquote {\bibinfo {title} {Effects of shell structure
  and {N}/{Z} ratio of a projectile on the excitation energy distribution
  between interacting nuclei in deep-inelastic collisions},}\ }\href {\doibase
  10.1103/physrevc.53.871} {\bibfield  {journal} {\bibinfo  {journal} {Phys.
  Rev. C}\ }\textbf {\bibinfo {volume} {53}},\ \bibinfo {pages} {871--879}
  (\bibinfo {year} {1996})}\BibitemShut {NoStop}%
\bibitem [{\citenamefont {Aritomo}(2006)}]{aritomo2006b}%
  \BibitemOpen
  \bibfield  {author} {\bibinfo {author} {\bibfnamefont {Y.}~\bibnamefont
  {Aritomo}},\ }\bibfield  {title} {\enquote {\bibinfo {title} {{F}usion
  hindrance and roles of shell effects in superheavy mass region},}\ }\href
  {\doibase 10.1016/j.nuclphysa.2006.09.018} {\bibfield  {journal} {\bibinfo
  {journal} {Nucl. Phys. A}\ }\textbf {\bibinfo {volume} {780}},\ \bibinfo
  {pages} {222} (\bibinfo {year} {2006})}\BibitemShut {NoStop}%
\bibitem [{\citenamefont {{Valery Zagrebaev}}\ and\ \citenamefont {{Walter
  Greiner}}(2007{\natexlab{b}})}]{zagrebaev2007}%
  \BibitemOpen
  \bibfield  {author} {\bibinfo {author} {\bibnamefont {{Valery Zagrebaev}}}\
  and\ \bibinfo {author} {\bibnamefont {{Walter Greiner}}},\ }\bibfield
  {title} {\enquote {\bibinfo {title} {{S}hell effects in damped collisions: a
  new way to superheavies},}\ }\href {\doibase 10.1088/0954-3899/34/11/004}
  {\bibfield  {journal} {\bibinfo  {journal} {J. Phys. G}\ }\textbf {\bibinfo
  {volume} {34}},\ \bibinfo {pages} {2265} (\bibinfo {year}
  {2007}{\natexlab{b}})}\BibitemShut {NoStop}%
\bibitem [{\citenamefont {Kozulin}\ \emph {et~al.}(2014)\citenamefont
  {Kozulin}, \citenamefont {Knyazheva}, \citenamefont {Itkis}, \citenamefont
  {Kozulina}, \citenamefont {Loktev}, \citenamefont {Novikov},\ and\
  \citenamefont {Harca}}]{kozulin2014b}%
  \BibitemOpen
  \bibfield  {author} {\bibinfo {author} {\bibfnamefont {E.~M.}\ \bibnamefont
  {Kozulin}}, \bibinfo {author} {\bibfnamefont {G.~N.}\ \bibnamefont
  {Knyazheva}}, \bibinfo {author} {\bibfnamefont {I.~M.}\ \bibnamefont
  {Itkis}}, \bibinfo {author} {\bibfnamefont {N.~I.}\ \bibnamefont {Kozulina}},
  \bibinfo {author} {\bibfnamefont {T.~A.}\ \bibnamefont {Loktev}}, \bibinfo
  {author} {\bibfnamefont {K.~V.}\ \bibnamefont {Novikov}}, \ and\ \bibinfo
  {author} {\bibfnamefont {I.}~\bibnamefont {Harca}},\ }\bibfield  {title}
  {\enquote {\bibinfo {title} {Shell effects in fission, quasifission and
  multinucleon transfer reaction},}\ }\href {\doibase
  10.1088/1742-6596/515/1/012010} {\bibfield  {journal} {\bibinfo  {journal}
  {J. Phys. Conf. Ser.}\ }\textbf {\bibinfo {volume} {515}},\ \bibinfo {pages}
  {012010} (\bibinfo {year} {2014})}\BibitemShut {NoStop}%
\bibitem [{\citenamefont {Mohanto}\ \emph {et~al.}(2018)\citenamefont
  {Mohanto}, \citenamefont {Hinde}, \citenamefont {Banerjee}, \citenamefont
  {Dasgupta}, \citenamefont {Jeung}, \citenamefont {Simenel}, \citenamefont
  {Simpson}, \citenamefont {Wakhle}, \citenamefont {Williams}, \citenamefont
  {Carter}, \citenamefont {Cook}, \citenamefont {Luong}, \citenamefont
  {Palshetkar},\ and\ \citenamefont {Rafferty}}]{mohanto2018}%
  \BibitemOpen
  \bibfield  {author} {\bibinfo {author} {\bibfnamefont {G.}~\bibnamefont
  {Mohanto}}, \bibinfo {author} {\bibfnamefont {D.~J.}\ \bibnamefont {Hinde}},
  \bibinfo {author} {\bibfnamefont {K.}~\bibnamefont {Banerjee}}, \bibinfo
  {author} {\bibfnamefont {M.}~\bibnamefont {Dasgupta}}, \bibinfo {author}
  {\bibfnamefont {D.~Y.}\ \bibnamefont {Jeung}}, \bibinfo {author}
  {\bibfnamefont {C.}~\bibnamefont {Simenel}}, \bibinfo {author} {\bibfnamefont
  {E.~C.}\ \bibnamefont {Simpson}}, \bibinfo {author} {\bibfnamefont
  {A.}~\bibnamefont {Wakhle}}, \bibinfo {author} {\bibfnamefont
  {E.}~\bibnamefont {Williams}}, \bibinfo {author} {\bibfnamefont {I.~P.}\
  \bibnamefont {Carter}}, \bibinfo {author} {\bibfnamefont {K.~J.}\
  \bibnamefont {Cook}}, \bibinfo {author} {\bibfnamefont {D.~H.}\ \bibnamefont
  {Luong}}, \bibinfo {author} {\bibfnamefont {C.~S.}\ \bibnamefont
  {Palshetkar}}, \ and\ \bibinfo {author} {\bibfnamefont {D.~C.}\ \bibnamefont
  {Rafferty}},\ }\bibfield  {title} {\enquote {\bibinfo {title} {Interplay of
  spherical closed shells and {$N/Z$} asymmetry in quasifission dynamics},}\
  }\href {\doibase 10.1103/PhysRevC.97.054603} {\bibfield  {journal} {\bibinfo
  {journal} {Phys. Rev. C}\ }\textbf {\bibinfo {volume} {97}},\ \bibinfo
  {pages} {054603} (\bibinfo {year} {2018})}\BibitemShut {NoStop}%
\bibitem [{\citenamefont {Umar}\ \emph {et~al.}(2015)\citenamefont {Umar},
  \citenamefont {Oberacker},\ and\ \citenamefont {Simenel}}]{umar2015a}%
  \BibitemOpen
  \bibfield  {author} {\bibinfo {author} {\bibfnamefont {A.~S.}\ \bibnamefont
  {Umar}}, \bibinfo {author} {\bibfnamefont {V.~E.}\ \bibnamefont {Oberacker}},
  \ and\ \bibinfo {author} {\bibfnamefont {C.}~\bibnamefont {Simenel}},\
  }\bibfield  {title} {\enquote {\bibinfo {title} {{S}hape evolution and
  collective dynamics of quasifission in the time-dependent {H}artree-{F}ock
  approach},}\ }\href {\doibase 10.1103/PhysRevC.92.024621} {\bibfield
  {journal} {\bibinfo  {journal} {Phys. Rev. C}\ }\textbf {\bibinfo {volume}
  {92}},\ \bibinfo {pages} {024621} (\bibinfo {year} {2015})}\BibitemShut
  {NoStop}%
\bibitem [{\citenamefont {Simenel}(2012)}]{simenel2012}%
  \BibitemOpen
  \bibfield  {author} {\bibinfo {author} {\bibfnamefont {C\'edric}\
  \bibnamefont {Simenel}},\ }\bibfield  {title} {\enquote {\bibinfo {title}
  {{N}uclear quantum many-body dynamics},}\ }\href {\doibase
  10.1140/epja/i2012-12152-0} {\bibfield  {journal} {\bibinfo  {journal} {Eur.
  Phys. J. A}\ }\textbf {\bibinfo {volume} {48}},\ \bibinfo {pages} {152}
  (\bibinfo {year} {2012})}\BibitemShut {NoStop}%
\bibitem [{\citenamefont {Simenel}\ and\ \citenamefont
  {Umar}(2018)}]{simenel2018}%
  \BibitemOpen
  \bibfield  {author} {\bibinfo {author} {\bibfnamefont {C.}~\bibnamefont
  {Simenel}}\ and\ \bibinfo {author} {\bibfnamefont {A.~S.}\ \bibnamefont
  {Umar}},\ }\bibfield  {title} {\enquote {\bibinfo {title} {Heavy--ion
  collisions and fission dynamics with the time--dependent {H}artree-{F}ock
  theory and its extensions},}\ }\href {https://arxiv.org/abs/1807.01859}
  {\bibfield  {journal} {\bibinfo  {journal} {arXiv:1807.01859}\ } (\bibinfo
  {year} {2018})}\BibitemShut {NoStop}%
\bibitem [{\citenamefont {Simenel}(2010)}]{simenel2010}%
  \BibitemOpen
  \bibfield  {author} {\bibinfo {author} {\bibfnamefont {C\'edric}\
  \bibnamefont {Simenel}},\ }\bibfield  {title} {\enquote {\bibinfo {title}
  {{P}article {T}ransfer {R}eactions with the {T}ime-{D}ependent
  {H}artree-{F}ock {T}heory {U}sing a {P}article {N}umber {P}rojection
  {T}echnique},}\ }\href {\doibase 10.1103/PhysRevLett.105.192701} {\bibfield
  {journal} {\bibinfo  {journal} {Phys. Rev. Lett.}\ }\textbf {\bibinfo
  {volume} {105}},\ \bibinfo {pages} {192701} (\bibinfo {year}
  {2010})}\BibitemShut {NoStop}%
\bibitem [{\citenamefont {Simenel}(2011)}]{simenel2011}%
  \BibitemOpen
  \bibfield  {author} {\bibinfo {author} {\bibfnamefont {C{\'e}dric}\
  \bibnamefont {Simenel}},\ }\bibfield  {title} {\enquote {\bibinfo {title}
  {{P}article-{N}umber {F}luctuations and {C}orrelations in {T}ransfer
  {R}eactions {O}btained {U}sing the {B}alian-{V}\'en\'eroni {V}ariational
  {P}rinciple},}\ }\href {\doibase 10.1103/PhysRevLett.106.112502} {\bibfield
  {journal} {\bibinfo  {journal} {Phys. Rev. Lett.}\ }\textbf {\bibinfo
  {volume} {106}},\ \bibinfo {pages} {112502} (\bibinfo {year}
  {2011})}\BibitemShut {NoStop}%
\bibitem [{\citenamefont {Scamps}\ and\ \citenamefont
  {Lacroix}(2013)}]{scamps2013a}%
  \BibitemOpen
  \bibfield  {author} {\bibinfo {author} {\bibfnamefont {Guillaume}\
  \bibnamefont {Scamps}}\ and\ \bibinfo {author} {\bibfnamefont {Denis}\
  \bibnamefont {Lacroix}},\ }\bibfield  {title} {\enquote {\bibinfo {title}
  {{E}ffect of pairing on one- and two-nucleon transfer below the {C}oulomb
  barrier: {A} time-dependent microscopic description},}\ }\href {\doibase
  10.1103/PhysRevC.87.014605} {\bibfield  {journal} {\bibinfo  {journal} {Phys.
  Rev. C}\ }\textbf {\bibinfo {volume} {87}},\ \bibinfo {pages} {014605}
  (\bibinfo {year} {2013})}\BibitemShut {NoStop}%
\bibitem [{\citenamefont {Scamps}\ \emph {et~al.}(2015)\citenamefont {Scamps},
  \citenamefont {Sargsyan}, \citenamefont {Adamian}, \citenamefont
  {Antonenko},\ and\ \citenamefont {Lacroix}}]{scamps2015d}%
  \BibitemOpen
  \bibfield  {author} {\bibinfo {author} {\bibfnamefont {G.}~\bibnamefont
  {Scamps}}, \bibinfo {author} {\bibfnamefont {V.~V.}\ \bibnamefont
  {Sargsyan}}, \bibinfo {author} {\bibfnamefont {G.~G.}\ \bibnamefont
  {Adamian}}, \bibinfo {author} {\bibfnamefont {N.~V.}\ \bibnamefont
  {Antonenko}}, \ and\ \bibinfo {author} {\bibfnamefont {D.}~\bibnamefont
  {Lacroix}},\ }\bibfield  {title} {\enquote {\bibinfo {title} {{A}nalysis of
  the dependence of the few-neutron transfer probability on the ${Q}$-value
  magnitudes},}\ }\href {\doibase 10.1103/PhysRevC.91.024601} {\bibfield
  {journal} {\bibinfo  {journal} {Phys. Rev. C}\ }\textbf {\bibinfo {volume}
  {91}},\ \bibinfo {pages} {024601} (\bibinfo {year} {2015})}\BibitemShut
  {NoStop}%
\bibitem [{\citenamefont {Scamps}\ and\ \citenamefont
  {Hashimoto}(2017)}]{scamps2017b}%
  \BibitemOpen
  \bibfield  {author} {\bibinfo {author} {\bibfnamefont {Guillaume}\
  \bibnamefont {Scamps}}\ and\ \bibinfo {author} {\bibfnamefont {Yukio}\
  \bibnamefont {Hashimoto}},\ }\bibfield  {title} {\enquote {\bibinfo {title}
  {Transfer probabilities for the reactions
  $^{14,20}\mathrm{O}+^{20}\mathrm{O}$ in terms of multiple time-dependent
  {H}artree-{F}ock-{B}ogoliubov trajectories},}\ }\href {\doibase
  10.1103/PhysRevC.96.031602} {\bibfield  {journal} {\bibinfo  {journal} {Phys.
  Rev. C}\ }\textbf {\bibinfo {volume} {96}},\ \bibinfo {pages} {031602}
  (\bibinfo {year} {2017})}\BibitemShut {NoStop}%
\bibitem [{\citenamefont {{Kazuyuki Sekizawa}}\ and\ \citenamefont {{Kazuhiro
  Yabana}}(2013)}]{sekizawa2013}%
  \BibitemOpen
  \bibfield  {author} {\bibinfo {author} {\bibnamefont {{Kazuyuki Sekizawa}}}\
  and\ \bibinfo {author} {\bibnamefont {{Kazuhiro Yabana}}},\ }\bibfield
  {title} {\enquote {\bibinfo {title} {{T}ime-dependent {H}artree-{F}ock
  calculations for multinucleon transfer processes in
  $^{40,48}${C}a+$^{124}${S}n, $^{40}${C}a+$^{208}${P}b, and
  $^{58}${N}i+$^{208}${P}b reactions},}\ }\href {\doibase
  10.1103/PhysRevC.88.014614} {\bibfield  {journal} {\bibinfo  {journal} {Phys.
  Rev. C}\ }\textbf {\bibinfo {volume} {88}},\ \bibinfo {pages} {014614}
  (\bibinfo {year} {2013})}\BibitemShut {NoStop}%
\bibitem [{\citenamefont {Sekizawa}\ and\ \citenamefont
  {Yabana}(2014)}]{sekizawa2014}%
  \BibitemOpen
  \bibfield  {author} {\bibinfo {author} {\bibfnamefont {Kazuyuki}\
  \bibnamefont {Sekizawa}}\ and\ \bibinfo {author} {\bibfnamefont {Kazuhiro}\
  \bibnamefont {Yabana}},\ }\bibfield  {title} {\enquote {\bibinfo {title}
  {{P}article-number projection method in time-dependent {H}artree-{F}ock
  theory: {P}roperties of reaction products},}\ }\href {\doibase
  10.1103/PhysRevC.90.064614} {\bibfield  {journal} {\bibinfo  {journal} {Phys.
  Rev. C}\ }\textbf {\bibinfo {volume} {90}},\ \bibinfo {pages} {064614}
  (\bibinfo {year} {2014})}\BibitemShut {NoStop}%
\bibitem [{\citenamefont {Sekizawa}\ and\ \citenamefont
  {Yabana}(2016)}]{sekizawa2016}%
  \BibitemOpen
  \bibfield  {author} {\bibinfo {author} {\bibfnamefont {Kazuyuki}\
  \bibnamefont {Sekizawa}}\ and\ \bibinfo {author} {\bibfnamefont {Kazuhiro}\
  \bibnamefont {Yabana}},\ }\bibfield  {title} {\enquote {\bibinfo {title}
  {{T}ime-dependent {H}artree-{F}ock calculations for multinucleon transfer and
  quasifission processes in the $^{64}\text{Ni}+^{238}\text{U}$ reaction},}\
  }\href {\doibase 10.1103/PhysRevC.93.054616} {\bibfield  {journal} {\bibinfo
  {journal} {Phys. Rev. C}\ }\textbf {\bibinfo {volume} {93}},\ \bibinfo
  {pages} {054616} (\bibinfo {year} {2016})}\BibitemShut {NoStop}%
\bibitem [{\citenamefont {Sekizawa}(2017)}]{sekizawa2017a}%
  \BibitemOpen
  \bibfield  {author} {\bibinfo {author} {\bibfnamefont {Kazuyuki}\
  \bibnamefont {Sekizawa}},\ }\bibfield  {title} {\enquote {\bibinfo {title}
  {Enhanced nucleon transfer in tip collisions of
  $^{238}\mathrm{U}+^{124}\mathrm{Sn}$},}\ }\href {\doibase
  10.1103/PhysRevC.96.041601} {\bibfield  {journal} {\bibinfo  {journal} {Phys.
  Rev. C}\ }\textbf {\bibinfo {volume} {96}},\ \bibinfo {pages} {041601(R)}
  (\bibinfo {year} {2017})}\BibitemShut {NoStop}%
\bibitem [{\citenamefont {Williams}\ \emph {et~al.}(2018)\citenamefont
  {Williams}, \citenamefont {Sekizawa}, \citenamefont {Hinde}, \citenamefont
  {Simenel}, \citenamefont {Dasgupta}, \citenamefont {Carter}, \citenamefont
  {Cook}, \citenamefont {Jeung}, \citenamefont {McNeil}, \citenamefont
  {Palshetkar}, \citenamefont {Rafferty}, \citenamefont {Ramachandran},\ and\
  \citenamefont {Wakhle}}]{williams2018}%
  \BibitemOpen
  \bibfield  {author} {\bibinfo {author} {\bibfnamefont {E.}~\bibnamefont
  {Williams}}, \bibinfo {author} {\bibfnamefont {K.}~\bibnamefont {Sekizawa}},
  \bibinfo {author} {\bibfnamefont {D.~J.}\ \bibnamefont {Hinde}}, \bibinfo
  {author} {\bibfnamefont {C.}~\bibnamefont {Simenel}}, \bibinfo {author}
  {\bibfnamefont {M.}~\bibnamefont {Dasgupta}}, \bibinfo {author}
  {\bibfnamefont {I.~P.}\ \bibnamefont {Carter}}, \bibinfo {author}
  {\bibfnamefont {K.~J.}\ \bibnamefont {Cook}}, \bibinfo {author}
  {\bibfnamefont {D.~Y.}\ \bibnamefont {Jeung}}, \bibinfo {author}
  {\bibfnamefont {S.~D.}\ \bibnamefont {McNeil}}, \bibinfo {author}
  {\bibfnamefont {C.~S.}\ \bibnamefont {Palshetkar}}, \bibinfo {author}
  {\bibfnamefont {D.~C.}\ \bibnamefont {Rafferty}}, \bibinfo {author}
  {\bibfnamefont {K.}~\bibnamefont {Ramachandran}}, \ and\ \bibinfo {author}
  {\bibfnamefont {A.}~\bibnamefont {Wakhle}},\ }\bibfield  {title} {\enquote
  {\bibinfo {title} {Exploring {Z}eptosecond {Q}uantum {E}quilibration
  {D}ynamics: {F}rom {D}eep-{I}nelastic to {F}usion-{F}ission {O}utcomes in
  $^{58}\mathrm{Ni}+^{60}\mathrm{Ni}$ {R}eactions},}\ }\href {\doibase
  10.1103/PhysRevLett.120.022501} {\bibfield  {journal} {\bibinfo  {journal}
  {Phys. Rev. Lett.}\ }\textbf {\bibinfo {volume} {120}},\ \bibinfo {pages}
  {022501} (\bibinfo {year} {2018})}\BibitemShut {NoStop}%
\bibitem [{\citenamefont {{Roger Balian}}\ and\ \citenamefont {{Marcel
  V\'en\'eroni}}(1984)}]{balian1984}%
  \BibitemOpen
  \bibfield  {author} {\bibinfo {author} {\bibnamefont {{Roger Balian}}}\ and\
  \bibinfo {author} {\bibnamefont {{Marcel V\'en\'eroni}}},\ }\bibfield
  {title} {\enquote {\bibinfo {title} {{F}luctuations in a time-dependent
  mean-field approach},}\ }\href {\doibase 10.1016/0370-2693(84)92008-2}
  {\bibfield  {journal} {\bibinfo  {journal} {Phys. Lett. B}\ }\textbf
  {\bibinfo {volume} {136}},\ \bibinfo {pages} {301--306} (\bibinfo {year}
  {1984})}\BibitemShut {NoStop}%
\bibitem [{\citenamefont {Broomfield}\ and\ \citenamefont
  {Stevenson}(2008)}]{broomfield2008}%
  \BibitemOpen
  \bibfield  {author} {\bibinfo {author} {\bibfnamefont {J.~M.~A.}\
  \bibnamefont {Broomfield}}\ and\ \bibinfo {author} {\bibfnamefont {P.~D.}\
  \bibnamefont {Stevenson}},\ }\bibfield  {title} {\enquote {\bibinfo {title}
  {{M}ass dispersions from giant dipole resonances using the
  {B}alian-{V}\'en\'eroni variational approach},}\ }\href {\doibase
  10.1088/0954-3899/35/9/095102} {\bibfield  {journal} {\bibinfo  {journal} {J.
  Phys. G}\ }\textbf {\bibinfo {volume} {35}},\ \bibinfo {pages} {095102}
  (\bibinfo {year} {2008})}\BibitemShut {NoStop}%
\bibitem [{\citenamefont {Ayik}(2008)}]{ayik2008}%
  \BibitemOpen
  \bibfield  {author} {\bibinfo {author} {\bibfnamefont {S.}~\bibnamefont
  {Ayik}},\ }\bibfield  {title} {\enquote {\bibinfo {title} {A stochastic
  mean-field approach for nuclear dynamics},}\ }\href {\doibase
  10.1016/j.physletb.2007.09.072} {\bibfield  {journal} {\bibinfo  {journal}
  {Phys. Lett. B}\ }\textbf {\bibinfo {volume} {658}},\ \bibinfo {pages} {174}
  (\bibinfo {year} {2008})}\BibitemShut {NoStop}%
\bibitem [{\citenamefont {Ayik}\ \emph {et~al.}(2017)\citenamefont {Ayik},
  \citenamefont {Yilmaz}, \citenamefont {Yilmaz}, \citenamefont {Umar},\ and\
  \citenamefont {Turan}}]{ayik2017}%
  \BibitemOpen
  \bibfield  {author} {\bibinfo {author} {\bibfnamefont {S.}~\bibnamefont
  {Ayik}}, \bibinfo {author} {\bibfnamefont {B.}~\bibnamefont {Yilmaz}},
  \bibinfo {author} {\bibfnamefont {O.}~\bibnamefont {Yilmaz}}, \bibinfo
  {author} {\bibfnamefont {A.~S.}\ \bibnamefont {Umar}}, \ and\ \bibinfo
  {author} {\bibfnamefont {G.}~\bibnamefont {Turan}},\ }\bibfield  {title}
  {\enquote {\bibinfo {title} {Multinucleon transfer in central collisions of
  $^{238}\mathrm{U}+^{238}\mathrm{U}$},}\ }\href {\doibase
  10.1103/PhysRevC.96.024611} {\bibfield  {journal} {\bibinfo  {journal} {Phys.
  Rev. C}\ }\textbf {\bibinfo {volume} {96}},\ \bibinfo {pages} {024611}
  (\bibinfo {year} {2017})}\BibitemShut {NoStop}%
\bibitem [{\citenamefont {Ayik}\ \emph {et~al.}(2018)\citenamefont {Ayik},
  \citenamefont {Yilmaz}, \citenamefont {Yilmaz},\ and\ \citenamefont
  {Umar}}]{ayik2018}%
  \BibitemOpen
  \bibfield  {author} {\bibinfo {author} {\bibfnamefont {S.}~\bibnamefont
  {Ayik}}, \bibinfo {author} {\bibfnamefont {B.}~\bibnamefont {Yilmaz}},
  \bibinfo {author} {\bibfnamefont {O.}~\bibnamefont {Yilmaz}}, \ and\ \bibinfo
  {author} {\bibfnamefont {A.~S.}\ \bibnamefont {Umar}},\ }\bibfield  {title}
  {\enquote {\bibinfo {title} {Quantal diffusion description of multinucleon
  transfers in heavy--ion collisions},}\ }\href {\doibase
  10.1103/PhysRevC.97.054618} {\bibfield  {journal} {\bibinfo  {journal} {Phys.
  Rev. C}\ }\textbf {\bibinfo {volume} {97}},\ \bibinfo {pages} {054618}
  (\bibinfo {year} {2018})}\BibitemShut {NoStop}%
\bibitem [{\citenamefont {Schr\"oder}\ \emph {et~al.}(1981)\citenamefont
  {Schr\"oder}, \citenamefont {Huizenga},\ and\ \citenamefont
  {Randrup}}]{schroder1981}%
  \BibitemOpen
  \bibfield  {author} {\bibinfo {author} {\bibfnamefont {W.~U.}\ \bibnamefont
  {Schr\"oder}}, \bibinfo {author} {\bibfnamefont {J.~R.}\ \bibnamefont
  {Huizenga}}, \ and\ \bibinfo {author} {\bibfnamefont {J.}~\bibnamefont
  {Randrup}},\ }\bibfield  {title} {\enquote {\bibinfo {title} {Correlated mass
  and charge transport induced by statistical nucleon exchange in damped
  nuclear reactions},}\ }\href {\doibase 10.1016/0370-2693(81)90924-2}
  {\bibfield  {journal} {\bibinfo  {journal} {Phys. Lett. B}\ }\textbf
  {\bibinfo {volume} {98}},\ \bibinfo {pages} {355--359} (\bibinfo {year}
  {1981})}\BibitemShut {NoStop}%
\bibitem [{\citenamefont {Merchant}\ and\ \citenamefont
  {N\"orenberg}(1981)}]{merchant1981}%
  \BibitemOpen
  \bibfield  {author} {\bibinfo {author} {\bibfnamefont {A.~C.}\ \bibnamefont
  {Merchant}}\ and\ \bibinfo {author} {\bibfnamefont {W.}~\bibnamefont
  {N\"orenberg}},\ }\bibfield  {title} {\enquote {\bibinfo {title} {Neutron and
  proton diffusion in heavy--ion collisions},}\ }\href {\doibase
  10.1016/0370-2693(81)90844-3} {\bibfield  {journal} {\bibinfo  {journal}
  {Phys. Lett. B}\ }\textbf {\bibinfo {volume} {104}},\ \bibinfo {pages}
  {15--18} (\bibinfo {year} {1981})}\BibitemShut {NoStop}%
\bibitem [{\citenamefont {Merchant}\ and\ \citenamefont
  {N\"orenberg}(1982)}]{merchant1982}%
  \BibitemOpen
  \bibfield  {author} {\bibinfo {author} {\bibfnamefont {A.~C.}\ \bibnamefont
  {Merchant}}\ and\ \bibinfo {author} {\bibfnamefont {W.}~\bibnamefont
  {N\"orenberg}},\ }\bibfield  {title} {\enquote {\bibinfo {title} {Microscopic
  transport theory of heavy-ion collisions},}\ }\href {\doibase
  10.1007/bf01415880} {\bibfield  {journal} {\bibinfo  {journal} {Z. Phys. A}\
  }\textbf {\bibinfo {volume} {308}},\ \bibinfo {pages} {315--327} (\bibinfo
  {year} {1982})}\BibitemShut {NoStop}%
\bibitem [{\citenamefont {Umar}\ \emph {et~al.}(1991)\citenamefont {Umar},
  \citenamefont {Strayer}, \citenamefont {Wu}, \citenamefont {Dean},\ and\
  \citenamefont {G\"u\c{c}l\"u}}]{umar1991a}%
  \BibitemOpen
  \bibfield  {author} {\bibinfo {author} {\bibfnamefont {A.~S.}\ \bibnamefont
  {Umar}}, \bibinfo {author} {\bibfnamefont {M.~R.}\ \bibnamefont {Strayer}},
  \bibinfo {author} {\bibfnamefont {J.~S.}\ \bibnamefont {Wu}}, \bibinfo
  {author} {\bibfnamefont {D.~J.}\ \bibnamefont {Dean}}, \ and\ \bibinfo
  {author} {\bibfnamefont {M.~C.}\ \bibnamefont {G\"u\c{c}l\"u}},\ }\bibfield
  {title} {\enquote {\bibinfo {title} {{N}uclear {H}artree-{F}ock calculations
  with splines},}\ }\href {\doibase 10.1103/PhysRevC.44.2512} {\bibfield
  {journal} {\bibinfo  {journal} {Phys. Rev. C}\ }\textbf {\bibinfo {volume}
  {44}},\ \bibinfo {pages} {2512--2521} (\bibinfo {year} {1991})}\BibitemShut
  {NoStop}%
\bibitem [{\citenamefont {Umar}\ and\ \citenamefont
  {Oberacker}(2006)}]{umar2006c}%
  \BibitemOpen
  \bibfield  {author} {\bibinfo {author} {\bibfnamefont {A.~S.}\ \bibnamefont
  {Umar}}\ and\ \bibinfo {author} {\bibfnamefont {V.~E.}\ \bibnamefont
  {Oberacker}},\ }\bibfield  {title} {\enquote {\bibinfo {title}
  {{T}hree-dimensional unrestricted time-dependent {H}artree-{F}ock fusion
  calculations using the full {S}kyrme interaction},}\ }\href {\doibase
  10.1103/PhysRevC.73.054607} {\bibfield  {journal} {\bibinfo  {journal} {Phys.
  Rev. C}\ }\textbf {\bibinfo {volume} {73}},\ \bibinfo {pages} {054607}
  (\bibinfo {year} {2006})}\BibitemShut {NoStop}%
\bibitem [{\citenamefont {{Ka--Hae Kim}}\ \emph {et~al.}(1997)\citenamefont
  {{Ka--Hae Kim}}, \citenamefont {{Takaharu Otsuka}},\ and\ \citenamefont
  {{Paul Bonche}}}]{kim1997}%
  \BibitemOpen
  \bibfield  {author} {\bibinfo {author} {\bibnamefont {{Ka--Hae Kim}}},
  \bibinfo {author} {\bibnamefont {{Takaharu Otsuka}}}, \ and\ \bibinfo
  {author} {\bibnamefont {{Paul Bonche}}},\ }\bibfield  {title} {\enquote
  {\bibinfo {title} {{T}hree-dimensional {TDHF} calculations for reactions of
  unstable nuclei},}\ }\href {\doibase 10.1088/0954-3899/23/10/014} {\bibfield
  {journal} {\bibinfo  {journal} {J. Phys. G}\ }\textbf {\bibinfo {volume}
  {23}},\ \bibinfo {pages} {1267} (\bibinfo {year} {1997})}\BibitemShut
  {NoStop}%
\bibitem [{\citenamefont {{Hannes Risken}}\ and\ \citenamefont {{Till
  Frank}}(1996)}]{risken1996}%
  \BibitemOpen
  \bibfield  {author} {\bibinfo {author} {\bibnamefont {{Hannes Risken}}}\ and\
  \bibinfo {author} {\bibnamefont {{Till Frank}}},\ }\href {\doibase
  10.1007/978-3-642-61544-3} {\emph {\bibinfo {title} {The {F}okker--{P}lanck
  {E}quation}}}\ (\bibinfo  {publisher} {Springer--Verlag},\ \bibinfo {address}
  {Berlin},\ \bibinfo {year} {1996})\BibitemShut {NoStop}%
\bibitem [{\citenamefont {Gardiner}(1991)}]{gardiner1991}%
  \BibitemOpen
  \bibfield  {author} {\bibinfo {author} {\bibfnamefont {C.~W.}\ \bibnamefont
  {Gardiner}},\ }\href@noop {} {\emph {\bibinfo {title} {Quantum {N}oise}}}\
  (\bibinfo  {publisher} {Springer--{V}erlag},\ \bibinfo {address} {Berlin},\
  \bibinfo {year} {1991})\BibitemShut {NoStop}%
\bibitem [{\citenamefont {Weiss}(1999)}]{weiss1999}%
  \BibitemOpen
  \bibfield  {author} {\bibinfo {author} {\bibfnamefont {U.}~\bibnamefont
  {Weiss}},\ }\href@noop {} {\emph {\bibinfo {title} {Quantum {D}issipative
  {S}ystems}}},\ \bibinfo {edition} {2nd}\ ed.\ (\bibinfo  {publisher} {World
  {S}cientific},\ \bibinfo {address} {Singapore},\ \bibinfo {year}
  {1999})\BibitemShut {NoStop}%
\end{thebibliography}%

\end{document}